\documentclass[useAMS,usenatbib,usegraphicx]{mn2e}

\usepackage{natbib}
\usepackage{epsfig}
\usepackage{amssymb}
\usepackage{amsfonts}
\usepackage{amsmath}
\usepackage{color}

\input epsf

\newcommand{\ltsimeq}{\raisebox{-0.6ex}{$\,\stackrel 
        {\raisebox{-.2ex}{$\textstyle <$}}{\sim}\,$}} 
\newcommand{\gtsimeq}{\raisebox{-0.6ex}{$\,\stackrel 
        {\raisebox{-.2ex}{$\textstyle >$}}{\sim}\,$}}

\begin{document}
\title[Determining neutrino properties using future galaxy redshift 
surveys]{Determining neutrino properties using future galaxy redshift surveys}
\author[]
{F. B. Abdalla$^{1,2}$,
S. Rawlings$^{1}$.
\\
$^{1}$Department of Physics, Oxford University,
Denys Wilkinson Building, Keble Road, Oxford OX1 3RH, U.K.\\
$^{2}$Department of Physics and Astronomy, University College London,
Gower Street, London, WC1E 6BT, UK.
}
\maketitle 

\begin{abstract}
Current measurements of the large-scale structure of galaxies are able to place an 
$\sim 0.5 ~ \rm eV$ upper limit on the absolute mass scale of neutrinos. An
order-of-magnitude improvement in raw sensitivity, together with an insensitivity to
systematic effects, is needed to reach the lowest value allowed by particle physics
experiments. We consider the prospects of determining both the neutrino mass scale 
and the number of of massive neutrinos using future redshift surveys, 
specifically those undertaken with the Square Kilometre Array (SKA), with and without
additional constraints from the upcoming Planck CMB experiment.
If the sum of the neutrino masses $\sum m_i \gtsimeq 0.25 ~ \rm eV$ then 
the imprint of neutrinos on large-scale structure (LSS) should be enough, on its own, to establish 
the neutrino mass scale and, considered alongside CMB constraints, it will also determine the
number of massive neutrinos $N_{\nu}$, and hence the mass hierarchy. 
If $\sum m_i \sim 0.05 ~ \rm eV$, at the bottom end of the allowed range, then 
a combination of LSS, CMB and particle physics constraints
should be able to determine $\sum m_i$, $N_{\nu}$ and the hierarchy. If 
$\sum m_i$ is in the specific range $0.1-0.25 ~ \rm eV$, then  
a combination of LSS, CMB and particle physics
experiments should determine $\sum m_i$, but not $N_{\nu}$ or the hierarchy.
Once an SKA-like LSS survey is available there are good prospects of obtaining a 
full understanding of the conventional neutrino sector, and a chance of finding evidence
for sterile neutrinos.
\end{abstract}


\section{Introduction}

The proof that neutrinos have mass 
was a major breakthrough in particle physics. 
This proof came from the observation \citep[e.g.][]{1996PhRvL..77.1683F} that neutrinos in
one weak-flavour state are able to `oscillate into' neutrinos of a 
different weak-flavour state which, in `the vacuum', is disallowed by quantum mechanics 
unless neutrinos have mass. The implications of this result are profound not only for 
particle physics, but also for cosmology. It was the first measured effect that is not included
in the Standard Model of particle physics. It also started to set constraints 
on the absolute mass scale of a particle species which cosmologists have yet to include properly
in the energy density budget of the Universe. A recent review of the critical issues 
is given by  \citet{Kayser} and another recent review approaching neutrino physics from the theoretical cosmology side is given in \citet{2006PhR...429..307L}.

Throughout this paper we keep with particle physics notation by using natural units in
which $c = \hbar = k = 1$. We will refer throughout to a `fiducial' cosmology in which
the normal cosmological constants take the values:
\{$\Omega_b$, $\Omega_c$, $h$, $n_s$, $\sigma_8$\} = \{0.04, 0.26, 0.72, 1.0, 0.9\}. 
We define $\Omega_{m}$ as the fraction of critical density contributed by all
matter: baryons, CDM and neutrinos. The power spectrum $P(k)$ is calculated using the `Boltzmann 
code' CAMB \citep{2000ApJ...538..473L}. We reserve the use of the symbol $m_{\nu}$ for the 
absolute mass scale of neutrinos, by which we mean the rest mass of the most massive 
neutrino (see Fig.~\ref{fig:neutrino_hyer}). 
Throughout this paper we assume that the Universe is flat and that the 
`dark energy' is Einstein's cosmological constant with an equation-of-state parameter
$w=-1$.

\subsection{Background Particle Physics}
\label{sec::back}

Vacuum neutrino oscillations occur if neutrinos are massive because
their mass eigenstates are an admixture of the weak-flavour eigenstates 
familiar from the physics of electroweak interactions. This can be written as

\begin{equation}
\left|\nu_\alpha\right>=\sum_i U_{\alpha i}^*\left|\nu_i\right>,
\label{eq::states_nu}
\end{equation}

\noindent where $\alpha = 1,2,3$ labels the weak-flavour eigenstates and $i = 1,2,3$ labels the
mass eigenstates. $U_{\alpha i}$ is the mixing matrix that governs the rate of 
oscillations between different weak-flavour eigenstates. This yields 
a probability of a given neutrino oscillating from a weak-flavour state $\alpha$ to another 
weak-flavour state $\beta$ of

\begin{equation}
P_{\alpha \beta}=|\left<\nu_\beta|\nu_\alpha(t)\right>|^2=
|\sum_i \sum_j U_{\alpha i}^* U_{\beta j} \left<\nu_j(0)|\nu_i(t)\right>|^2.
\label{eq::prob_states_nu}
\end{equation}

\begin{figure}
\begin{center}
\centerline{
\includegraphics[width=9.0cm,angle=0]{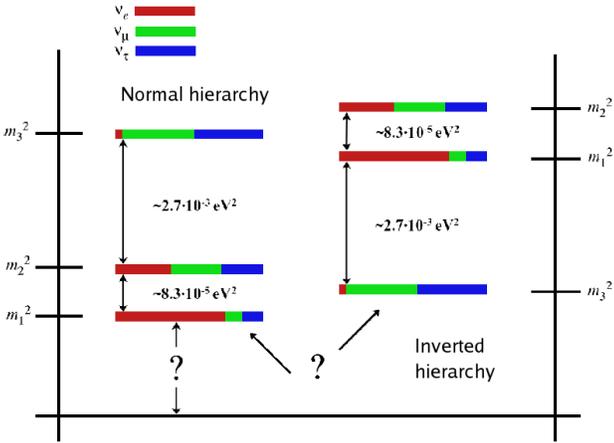}}
\caption[Neutrino hierarchies.]{ The (mass)$^{2}$ spectrum of neutrinos 
allowed by current neutrino oscillation experiments for (left) a `normal'
hierarchy and (right) an `inverted' hierarchy. We define
the absolute mass scale of the neutrino $m_{\nu}$ as the mass of the
most massive eigenstate. This scale is bounded below by  
the minimum required by oscillation experiments ($\sim 0.05 ~ \rm  eV$) 
and above (at $\sim 0.5 \, {\rm eV}$) by current cosmological (large-scale structure) measurements
\citep[e.g.][]{2005NJPh....7...61E,2005PhRvD..71j3515S,2006PhRvD..74b7302F}, some of which are expected to have more systematic effects than others. If the true value of $m_{\nu}$ lies towards the 
top of this allowed range, then we call this a quasi-degenerate scenario because, 
regardless of the hierarchy, $m_{1} \approx m_{2} \approx m_{3}$
(with $m_{\nu}=m_{3}$ for a normal hierarchy, and $m_{\nu}=m_{2}$ for
an inverted hierarchy). 
If the true value of $m_\nu$
is towards the bottom of the allowed range $m_\nu = m_3$
(with $m_1 \approx m_2 \approx 0$) for a normal hierarchy; 
and $m_\nu = m_2 \approx m_1$ (with $m_3 \approx 0$)
 for an inverted hierarchy. The size of each coloured
bar is proportional to the contribution of weak-flavour 
eigenstates to each mass
eigenstate: red, $\nu_{\rm e} [|U_{ {\rm e} i}|^{2}]$;
green, $\nu_{\rm \nu} [|U_{ {\rm \nu} i}|^{2}]$;
and blue, $\nu_{\rm \tau} [|U_{ {\rm \tau} i}|^{2}]$.
\label{fig:neutrino_hyer}}
\end{center}
\end{figure}

The relationships between the weak-flavour and the mass eigenvectors for
neutrinos are illustrated in Fig.~\ref{fig:neutrino_hyer}. The third mass eigenstate 
is almost perpendicular to the axis of the electron neutrino eigenstate, or to put it in another way
the `mixing angle' defined by $\sin^{2} \theta_{13} = |U_{\rm e 3}|^{2}$ is small.
The experimental evidence for oscillations is summarised by \citet{Kayser}, but can be very
crudely summarised as follows: experiments that probe muon and electron anti-neutrinos 
originating from cosmic ray showers in the atmosphere can be approximated 
by a two-neutrino oscillation between muon and tau anti-neutrinos, controlled largely by 
$\Delta m_{32}^2$; experiments that probe electron anti-neutrinos originating in the sun find 
strong evidence for a different, but again approximately two-neutrino, oscillation between
electron and muon neutrinos, controlled largely by $\Delta m_{12}^2$. 
The current status of quantitative observational results are summarised by \citet{2004NJPh....6..122M}
(see also Fig.~\ref{fig:neutrino_hyer}).

The mixing matrix can be written in terms of a combination of Euler
rotations in an $n=3$ dimensional space, namely

\begin{eqnarray}
U_{\alpha i}=
\begin{pmatrix}
1&0&0 \cr 0& {c_{23}} & {s_{23}} \cr
0& -{s_{23}}& {c_{23}}
\end{pmatrix}
\times
\begin{pmatrix} 
{c_{13}} & 0 & {s_{13}}e^{i {\delta}}\cr
0&1&0\cr -{ s_{13}}e^{-i {\delta}} & 0  & 
{c_{13}}\cr
\end{pmatrix}
\nonumber
\\
\times 
\begin{pmatrix} 
{c_{21}} & {s_{12}}&0\cr
-{s_{12}}& {c_{12}}&0\cr
0&0&1
\end{pmatrix},
\label{eq::mix_mat}
\end{eqnarray}

\noindent where $c_{ij} = {\rm cos} \theta_{ij}$ and $s_{ij}={\rm sin} \theta_{ij}$
are functions of the `mixing angles' $\theta_{ij}$.
We have assumed that neutrinos behave like Dirac particles  --
so that each neutrino and its associated anti-neutrino are distinguishable --
and we have introduced a phase $\delta$ which
allows for Charge-Parity (CP) violations in the sense of small differences between the 
probability of oscillations between neutrinos and `CP-mirror-image' oscillations between 
anti-neutrinos.

The transition probability can now be written as

\begin{equation}
{ P_{\alpha\beta}} ={\delta_{\alpha\beta}-4\sum_{i \ne j}\sum_{j>i}}
{ \mbox{Re}[U_{\alpha i}U^*_{\beta i} U^*_{\alpha j} U_{\beta j}]} 
\left({\rm sin}^2 \frac{L\Delta m_{ij}^2}{4E} \right)\; ,  
\label{eq::p_ab}
\end{equation}

\noindent where $L$ is the distance between the source and the observer, which is
assumed to be a vacuum, and $\Delta m_{ij}^2 =m^2_i-m^2_j$ because vacuum oscillations
conserve energy, $E = E_\alpha = E_\beta $. From Eqn.~\ref{eq::p_ab} it is clear why 
particle physics experiments are mainly sensitive to the differences between the 
squares of the masses of the mass eigenstates and the mixing angles.

The three main goals of future research into the neutrino sector are as follows: 
(i) to determine the absolute mass scale for neutrinos; (ii) to determine whether 
the hierarchy is normal or inverted (see Fig.~\ref{fig:neutrino_hyer}); 
(iii) to check whether there are so-called sterile neutrinos, particles which interact
only via gravity, additional to the three weak-flavour eigenstates. 
The prospects of rapid progress via further particle physics 
experiments is limited \citep{Kayser} and strongly dependent 
on unknown factors like the true magnitude of $\theta_{13}$, and whether or not the 
extra diagnostic power available, in principle, from measuring oscillations in the
presence of matter can be successfully harnessed.

\subsection{Background Cosmology}

Neutrinos decoupled from radiation and matter
within the first few minutes of the history of the Universe.
As this de-coupling occurred before electron-positron annihilation produced
an extra source of entropy for the photons, the neutrino temperature
$T_\nu$ has remained at a fraction of the temperature $T_\gamma$ of the CMB given by

\begin{equation}
T_\nu = \left(\frac{4}{11}\right)^{1/3} T_\gamma.
\end{equation}

Neutrinos remain relativistic until the point at which $T_\nu \simeq m_\nu$, which 
corresponds to a redshift $z \sim 6000 $ for $m_\nu \sim 1 ~ \rm eV$
or $z \sim 30$ for $m_\nu \sim 0.05 ~ \rm eV$. This means that neutrino free streaming
can suppress structure formation on small scales at early times, but the 
neutrino energy density is dominated by the rest mass at recent times. We
assume that there are $N_\nu$ `massive' neutrinos where $N_\nu \sim 3$ for a
quasi-degenerate scenario, $N_\nu \sim 2$ for an inverted, non-quasi-degenerate scenario
and $N_\nu \sim 1$ for a normal, non-quasi-degenerate scenario
(see Fig.~\ref{fig:neutrino_hyer}). We further assume that neutrinos are
Dirac particles so that each neutrino and its associated anti-neutrino are distinguishable.
We can then relate the energy density of photons (bosons) to the energy density of 
neutrinos (fermions). The number density of neutrinos can be related to the number 
density of photons by $n_\nu=(3/11) n_\gamma N_\nu$, so we can write

\begin{equation}
\Omega_{\nu} = 
\frac{\Sigma_i m_i}{93.14 h^2 ~ \rm eV} \simeq
\frac{3}{11}
\frac{m_\nu N_\nu n_\gamma}{\rho_c} \simeq
\frac{m_\nu N_\nu}{93.14 h^2 ~ \rm eV} 
 .
\label{eq::omega_nu_nb}
\end{equation}

Neutrinos behave as a component that transitions from having an
equation-of-state $p = w_\nu \rho$, where $p$ and $\rho$ are the
pressure and energy density respectively, with $w_\nu = 1/3$ 
to $w_\nu \rightarrow 0$ as they become non-relativistic. We can evaluate
$\omega_\nu(z)$ for neutrinos using

\begin{equation}
\begin{array}{l}
\rho =\displaystyle{g\int E(p)f(x,p)\frac{{\rm d}^3p}{(2\pi)^3}}\\
P = g \displaystyle{\int \frac{p^2}{3E(p)}f(x,p)\frac{{\rm d}^3p}{(2\pi)^3}},
\end{array}
\label{eq::p_rho_nu}
\end{equation}

\noindent where 
$g$ is the degeneracy and $f$ is the phase-space distribution function for 
fermions.

\begin{figure}
\begin{center}
\centerline{
\includegraphics[width=9.0cm,angle=0]{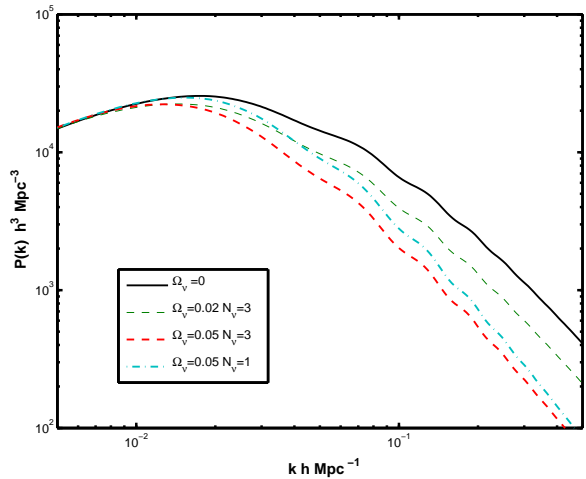}}
\caption[$P(k)$ for fiducial cosmology with the addition of neutrino parameters]{The 
power spectrum $P(k)$ for our fiducial cosmology (solid black line)
compared with those with different values of $\Omega_\nu$ and $N_\nu$, and through
Eqn.\ref{eq::omega_nu_nb}, $m_\nu$ (coloured dashed lines). There is a clear damping length scale imprinted
on $P(k)$ due to the free-streaming of neutrinos at early times
(Eqn.~\ref{eq::k_nu_lenght}). This scale is obviously dependent on 
$\Omega_\nu$ (Eqn.~\ref{eq::nu_pk}) and more subtly on $N_{\nu}$ because, 
for a given $\Omega_\nu$, a smaller $N_{\nu}$ delivers a larger $m_{\nu}$
which goes non-relativistic earlier and hence produces less damping.
LSS data should hence be able to constrain both the mass and the number of 
massive neutrino eigenstates.
\label{fig::t_k_neutrino}}
\end{center}
\end{figure}

The effect of neutrinos on large-scale structure is illustrated in 
Fig.~\ref{fig::t_k_neutrino}. The main feature is a strong attenuation of the 
power spectrum $P(k)$ on small scales if there is a significant energy density of neutrinos. 
This is associated with a preferred length (and $k$) scale given by the horizon when the
neutrinos become non-relativistic which, from Eqn.~1 of \citep{1998PhRvL..80.5255H}, is

\begin{equation}
k_{\nu} \simeq 0.026 \left( \frac{m_\nu}{\rm 1 ~ eV} \right)^{1/2} 
\Omega_m^{1/2} ~ h ~ {\rm Mpc}^{-1},
\label{eq::k_nu_lenght}
\end{equation}

\noindent assuming a quasi-degenerate scenario,
where $k_{\nu}$ is the associated co-moving wavenumber.
The damping of the power spectrum at large scales, small $k$, 
is \citep[Eqn.~2 of ][]{1998PhRvL..80.5255H} given approximately by

\begin{equation}
\frac{\Delta P(k)}{P(k)} \simeq -8 \frac{\Omega_{\nu}}{\Omega_{m}} 
                         \simeq -8 \left( \frac{m_\nu}{93.14 ~ \rm eV} \right)
                                     \left( \frac{N_\nu}{\Omega_{m} ~ h^{2}} \right).
\label{eq::nu_pk}
\end{equation}

\noindent This approximation, although not necessarily accurate on
all scales, can be used pedagogically in the following way.  
This expression assumes $\Omega_\nu \ll \Omega_m$ which is reasonable 
because current limits on neutrino energy density 
\citep[e.g.][]{2002PhRvL..89f1301E}
firmly exclude anything close to a Hot Dark Matter cosmological model.
To measure the effect that neutrinos of a given $m_\nu$
produce we need to measure $P(k)$ more accurately 
than the fractional shift given by Eqn.\ref{eq::nu_pk},
and we need to ensure that any systematic errors can be neglected.

It is possible to constrain the mass of the neutrino with CMB experiments. 
However, there are strong parameter degeneracies that prevent this being 
a very precise tool. It has been shown, nevertheless, that
even a neutrino mass as small as 0.05 eV could have a significant effect
on the CMB temperature and polarisation power 
spectrum \citep{1998PhRvL..80.5255H,2003PhRvD..67h5017H}.

\begin{figure*}
\begin{center}

\begin{minipage}[c]{.75\textwidth} 
\centering 
\includegraphics[width=9.5cm,angle=0]{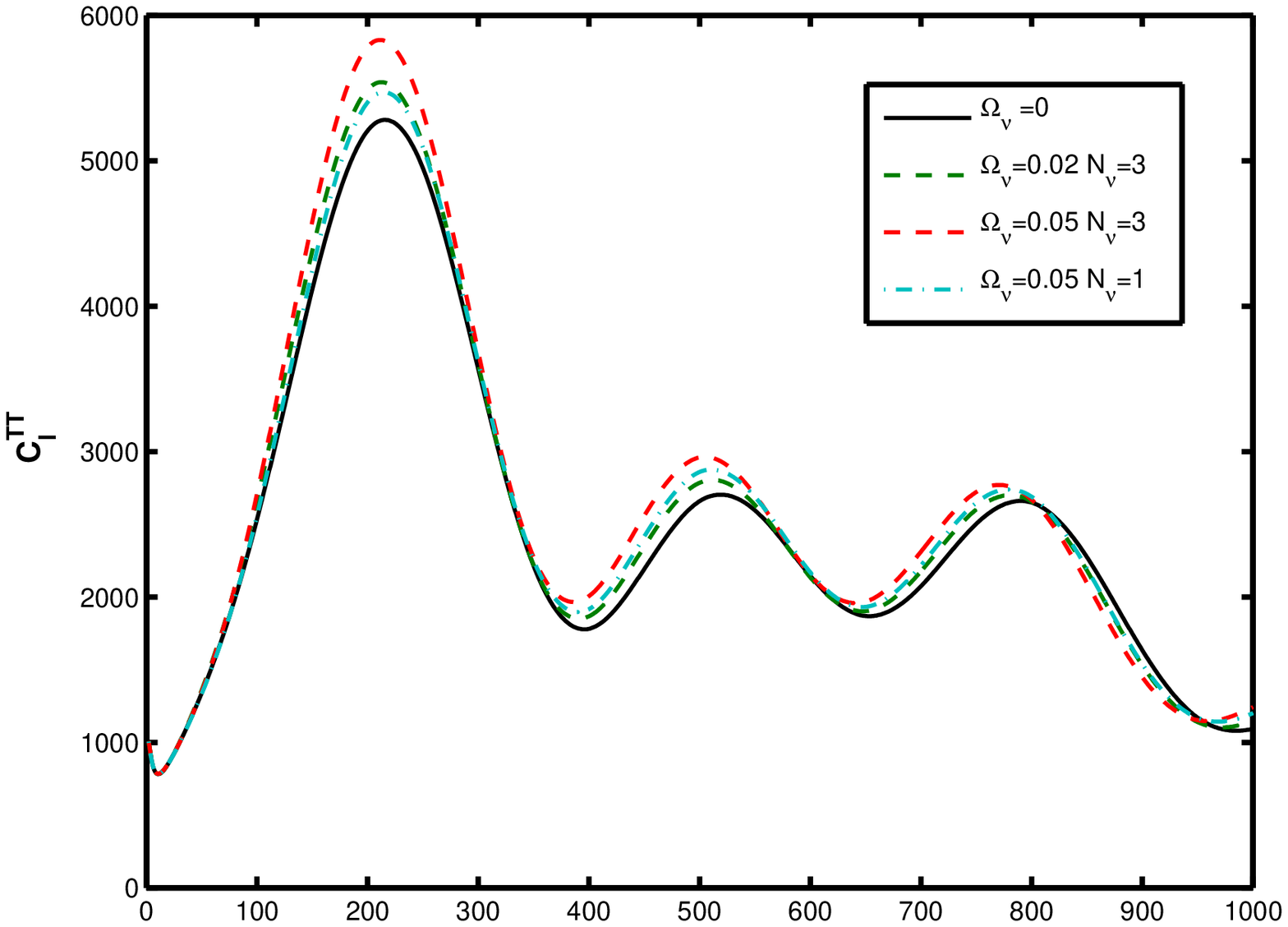}
\end{minipage}
\centerline{
\includegraphics[width=7.5cm,angle=0]{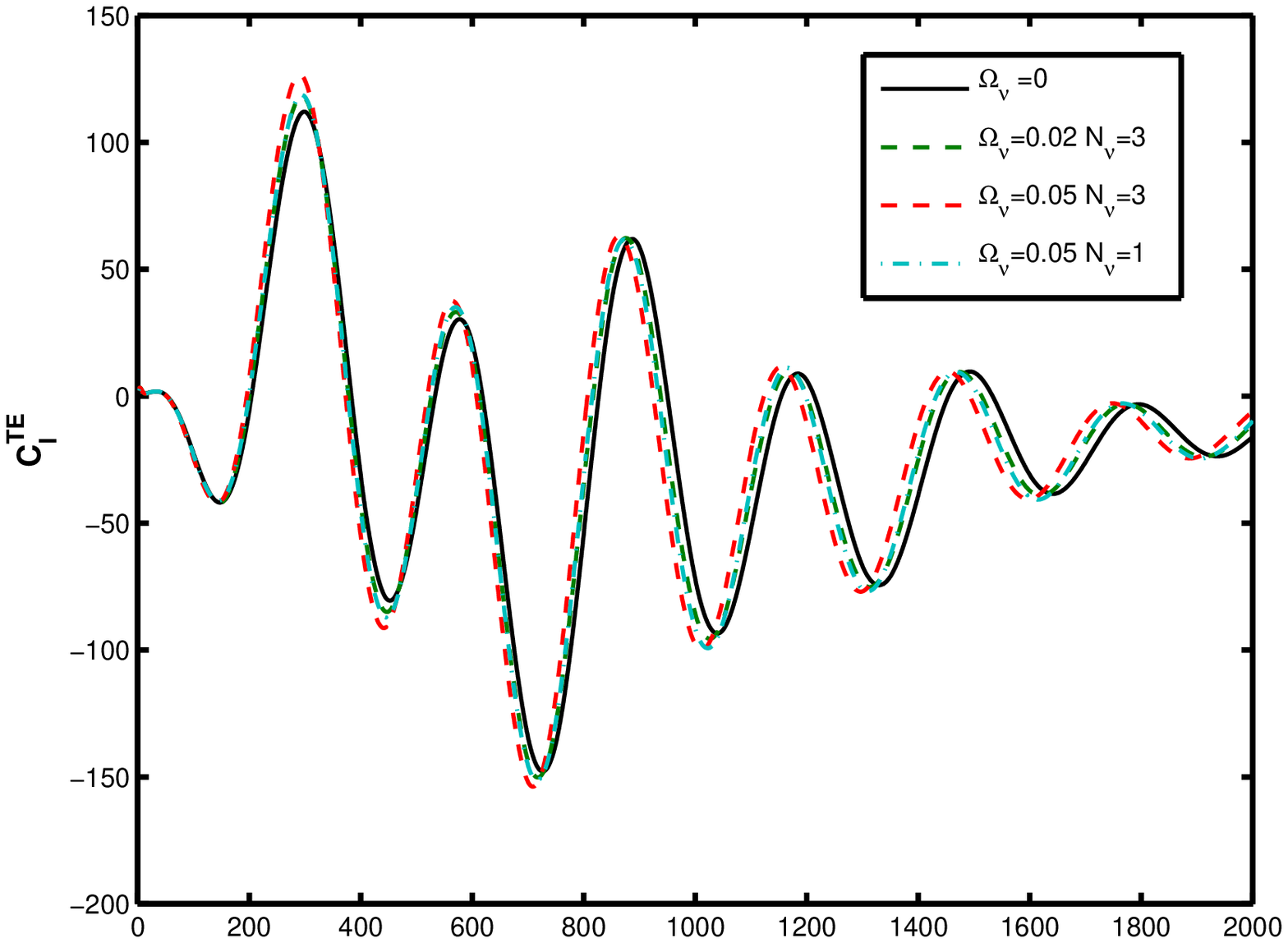}
\includegraphics[width=7.5cm,angle=0]{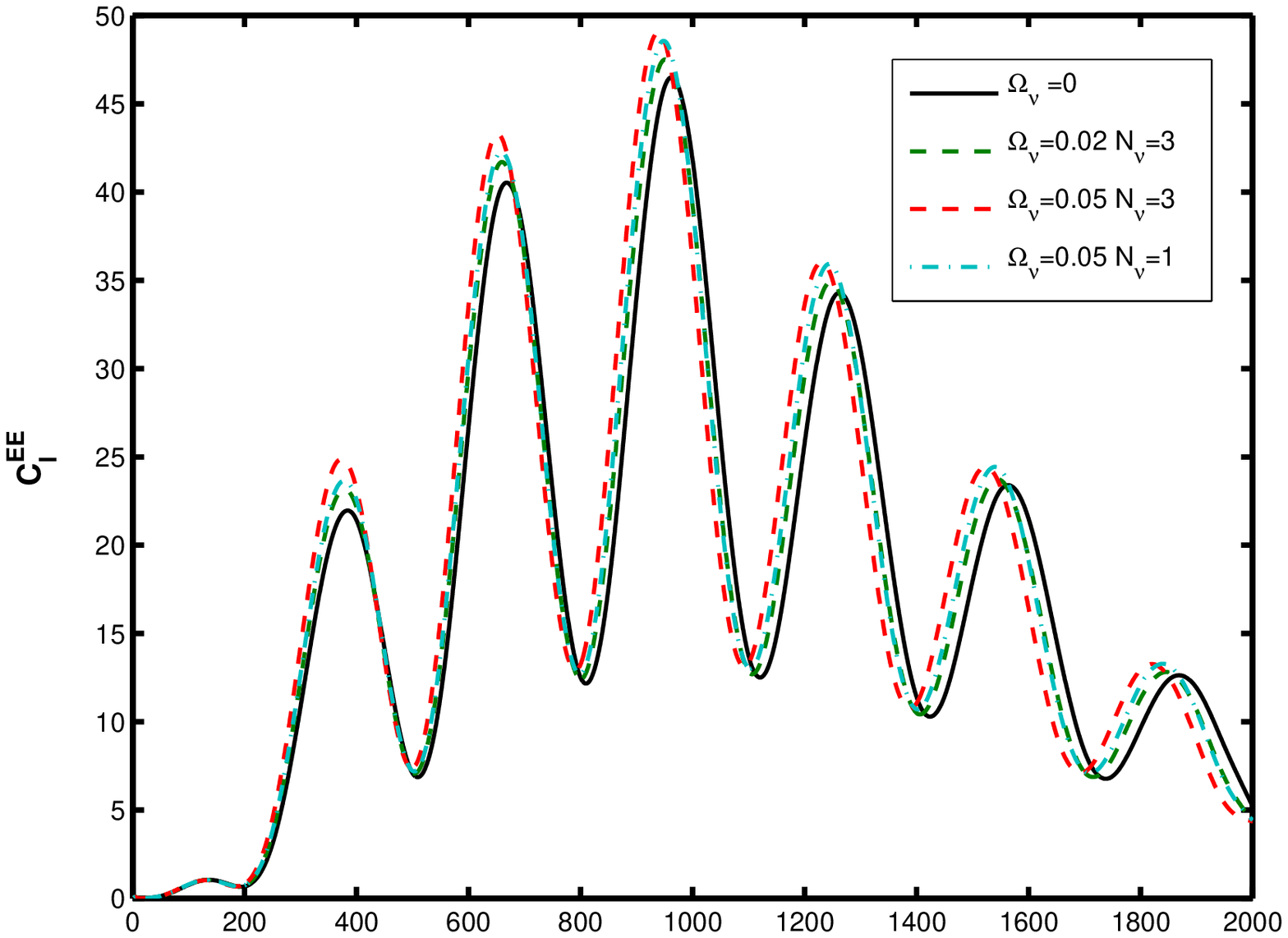}
}

\caption[Temperature and polarisation power spectra for a given cosmology:  
fiducial plus \{$N_\nu$, $\Omega_\nu$\}.]{The CMB temperature and
polarisation
power spectra for our fiducial cosmology (solid black line)
compared with those with different values of $\Omega_\nu$ and $N_\nu$, 
and through
Eqn.\ref{eq::omega_nu_nb}, $m_\nu$ (coloured dashed lines). 
Broadly speaking, when neutrinos are added, there is an increase 
in the total mass density (at the expense of dark energy density),
a change in the time of matter-radiation equality and a 
change in the amount of radiation at times before the neutrinos became cold. This
produces the changes seen in this figure.
There is a clear signal imprinted
on the power spectra due to the free-streaming of neutrinos at early times
(Eqn.~\ref{eq::k_nu_lenght}). This is dependent on 
$\Omega_\nu$ (Eqn.~\ref{eq::nu_pk}) and more subtly on $N_{\nu}$, however
as we will see degeneracies in the CMB analysis make it impossible to
disentangle the real values of \{$N_\nu$, $\Omega_\nu$\} with CMB data alone.
\label{fig::cl_neutrino}}
\end{center}
\end{figure*}

We can see in both Fig.\ref{fig::t_k_neutrino} and Fig.\ref{fig::cl_neutrino}
that adding two extra parameters \{$\Omega_\nu$, $N_\nu$\} to our fiducial
cosmology yields different fluctuations for different values of the parameters.
The total energy density in neutrinos $\Omega_\nu$ is the most important
quantity influencing $P(k)$ (Eqn.\ref{eq::nu_pk}) but, as explained
in the caption to Fig.~\ref{fig::t_k_neutrino}, the 
suppression of power also depends on $N_\nu$. 
CMB datasets are prone to serious
problems with parameter degeneracies: the 
bound on the neutrino mass are
\citep{2005PhRvD..71d3001I} claimed to be 
$\sum m_i < 2.0\, {\rm eV}$
from WMAP data alone.

\section{Methods of making forecasts for future cosmology experiments}

Cosmological experiments hold a lot of promise for
helping to address the key unknowns in the physics of the neutrino sector.
In this section we produce forecasts for the progress possible with 
upcoming CMB and LSS experiments.

To make these forecasts we will use Markov Chain Monte Carlo (MCMC)
methods to probe the $n-$dimensional parameter space of cosmological models with different
neutrino properties. Such methods \citep[e.g.][]{2002PhRvD..66b3528B} provide a powerful way of
forecasting the posterior probability distributions for a given set of cosmological
parameters to be probed by a future experiment. 

A MCMC is a chain of points drawn and chosen/rejected from a 
given distribution according to 
a certain number of rules. Points are chosen from a proposal 
distribution and either accepted or rejected according to the likelihood 
attached to both the chosen point and the previous point only (hence a Markov chain). In the simplest algorithms, if the new point has
a larger likelihood it is accepted and if the likelihood is lower it
is accepted with a probability which equals the ratio of the likelihoods 
from the old point and the new proposal point. The method is described fully by \citet{2002PhRvD..66j3511L}. If so, the density of points will describe the likelihood function in parameter space.

Given that no real data are available, we proceed 
in the following way. We first adopt a fiducial cosmology which we will assume 
is close enough to the truth that it can be used to deliver simulated data. 
We adopt here a theoretical framework used in \citet{2002PhRvD..66b3528B}; i.e.
we do not produce ensembles of mock datasets 
but assume that the simulated data $data_{sim}$ is precisely equivalent to 
the prediction of the fiducial model with an associated, appropriately
calculated, error bar. Any single realisation of the data would displace the
error ellipse found but would not change its size considerably.
The sum over many realisations of the data would produce results which are
equivalent to the results obtained by taking the data being equal to the
fiducial model, provided, of course, the fiducial model
is correct. We use this method for both LSS and CMB forecasts.
We then attach theoretical
errors to each fiducial data point in line with the errors expected
for the future cosmology experiment. Then, we used 
MCMC chains to map out the $n-$dimensional posterior 
probability distribution function $P(\theta_i|data_{sim})$ , where

\begin{equation}
P(\theta_i|data_{sim})  \propto P(data_{sim} |\theta_i) \times P(\theta_i),
\end{equation}

\noindent and $P(data_{sim} |\theta_i)$ is the likelihood function 
(given by $exp(-\chi^{2}/2)$ in the case of galaxy surveys
where $\chi^{2}$ is the usual chi squared, i.e. the sum over all simulated
data points of the square of the
data minus the model divided by the error estimate at that data point) 
and $P(\theta_i)$ is the $n$-dimensional
prior probability distribution function. 
We used our fiducial $\Lambda CDM$ model with the addition of the
neutrino parameters $\Omega_{\nu}$ and $N_{\nu}$ meaning $n=7$.
In the case of simulated LSS surveys
more `nuisance parameters' were needed to account for the way in which galaxies trace
dark matter: two per redshift shell, making a total of 13 parameters. In the case of the CMB
we added the optical depth of reionisation giving a total of 8 parameters.
We checked each MCMC chain had converged in two ways: we checked that the means and 
variances for each of the individual chains (eight MCMC chains were run for each case) 
were consistent with the other chains and we made a power spectrum 
amalysis of the chains \citep{2005MNRAS.356..925D}. Finally, we used the returned MCMC
samples to map out the posterior probability distribution.

Our methods are very similar to the Fisher
matrix approach \citep[e.g.][]{2003PhRvD..68f3004H,2003ApJ...598..720S}, 
in which one also assumes a fiducial cosmology and then
calculates the curvature of the posterior probability, assumed to be an 
$n-$dimensional Gaussian, around that point.
Both methods assume that the likelihood posterior is a slowly varying
function of the fiducial cosmology, but our approach has the advantage of 
directly mapping out any non-Gaussian features of the posterior 
probability distribution; in the case of a Fisher analysis
any non-Gaussianities of the posterior distribution function
may corrupt the error analysis.

\subsection{Methods of making forecasts for future LSS experiments}

It has been proposed that probing the pattern
of matter fluctuations as a function of redshift yields measurements
of the Hubble `Constant' $H(z)$ and the comoving distance $D(z)$ via a characteristic 
co-moving length scale imprinted on the power spectrum $P(k)$ via baryonic oscillations 
\citep{2003ApJ...594..665B,2003PhRvD..68f3004H,2003ApJ...598..720S}.
This characteristic co-moving length is derivable from known physics, being
essentially the size of the sound horizon at the time of last scattering. Given that 
matter and radiation dominate over dark energy at that period, its length is given by

\begin{equation}
r_s=\frac{c}{H_0}\int_0^{a_{rec}} \frac{1}{(3(1+R))^{1/2}} \frac{{\rm d}a}{a E(a)},
\end{equation}

\noindent where $R = 3 \rho_b / 4 \rho_\gamma$ introduces
a weak dependence of the sound speed of the plasma on
$\Omega_b$ and $a_{rec}$ is the scale factor at the redshift of recombination. 
We can interpret this as a sound speed
that is equal to $c/\sqrt{3}$ in the case of radiation only and which
decreases as a larger fraction of baryons is present.

Therefore we know the precise length of this standard rod for a given 
cosmology. Furthermore, this distance is dependent on $h$, $\Omega_m$ and
weakly on $\Omega_b$ but it is very insensitive to $w$ given that this acoustic scale was
set up in the early universe when it is thought that dark energy plays a small role.

A related geometric probe of the evolution of the 
Universe was proposed by \citet{1979Natur.281..358A}
by assuming that any characteristic length present in the Universe,
and which can be seen in both radial and angular directions in the sky,
produces a measurement of the product $H(z)D_A(z)$ [where 
$D_A(z) = D(z)/(1+z)$ is the angular diameter distance] 
simply because, by isotropy, 
the length has to be the same in both directions.

This proposed wiggles test 
\citep[e.g][]{2003ApJ...594..665B,2003PhRvD..68f3004H,2003ApJ...598..720S} 
can therefore simply be considered as a more powerful version of the 
Alcock-Paczynski (AP)
test and should be more useful in constraining dark energy as a geometrical
test which is weakly dependent on $w$.

More precisely, the comoving sizes of any object or any 
feature has a transverse ($r_{\perp}$) and 
parallel $r_{\parallel}$ projection on the sky which can be used for this
kind of test. These comoving features 
relate to the sizes of the observed
angular and redshift distances $\Delta \theta$ and $\Delta z$ respectively
via the following relations

\begin{equation}
\begin{array}{l}
\displaystyle{r_{\parallel}= \frac{c \Delta z}{H(z)}} \\
r_{\perp}=(1+z) D_A(z) \Delta \theta.
\end{array}
\end{equation}

Hence, when the true scales are known the measurement of the pattern in $z$ or
the line of sight
and $\theta$ or in the plane perpendicular to the line of sight
yield a measurement of $H(z)$ and $D_A(z)$, whereas
if the precise value of $r$ is not known, it is only the product 
$H(z) D_A(z)$ that can be measured.

To be able to estimate the errors on cosmological parameters
we have to know how well we will be able to measure the power spectrum. 
Under Gaussian conditions, the fractional error on the power spectrum
will depend on the total volume of the survey as this will determine the
number of Fourier modes that will be sampled by the survey;
the associated error is usually called cosmic variance. Furthermore
there will be an uncertainty given by the finite number of galaxies.
This is commonly termed shot noise. Hence the fractional error on the power
spectrum can be written as \citep{1994ApJ...426...23F}

\begin{equation}
\left(\frac{\Delta P}{P}\right)^2= 
\frac{4\pi}{V k^2\Delta k \Delta \mu}\left(\frac{1+nP}{nP}\right)^2
\label{eq::error_pk}
\end{equation}

\noindent where $P$ is the power spectrum measured, $n$ is the comoving number density of galaxies probed and $\mu$ is the cosine of the angle defined by our line of sight and the line joining pairs of galaxies in 3D space.

If we make a measurement of $P(k)$ several effects
can change, not only the shape, but also the
height of the power spectrum, as a function of redshift. 
If a wrong cosmology is assumed our estimates of distances 
and volumes are incorrect.
This produces a distortion that is visible in the power spectrum
as rings in the (perpendicular, parallel) $k$ plane
\citep{2003PhRvD..68f3004H}. Furthermore,
at a higher redshift, the growth factor changes the height of the
power spectrum. However if a wrong cosmology is assumed, then
we have wrong measurements of the cosmological volume in which we made
our survey and therefore the height of the power spectrum is measured 
incorrectly. These distortion effects have been thoroughly described in
\citet{1996MNRAS.282..877B}.

If the galaxies are measured in redshift and we do not have any information
about their peculiar velocities, we then retrieve 
a redshift space power spectrum.
This will differ from the real space power spectrum in two ways, 
firstly we obtain more 
correlations at low $k$ because large-scale bulk flows point 
directly towards matter overdensities due to gravitational pull. 
Therefore there is an enhancement factor
$(1+f_{k}\mu^2)^2$ derived in
\citet{1987MNRAS.227....1K}, where $\mu$ is the
cosine of the angle that the mode which is probed makes with the line
of sight, and $f_{k}$ is the derivative of the natural log of the 
over-density with respect to the natural log of the scale factor
$f_{k}={\rm d ln}\delta /{\rm d ln}a$.
At large scales,
structures appear to be closer to each other in redshift space. On the other
hand at small scales, given the circular velocities of satellite and
relaxed structures, redshift-space structures appear to be 
elongated along the line-of-sight. This created the so 
called finger-of-god effect \citep{2001Natur.410..169P}. Here
we are interested in the large-scale fluctuations, and we simply ignore
the small-scale power , i.e. we include the first effect
but not the finger-of-god effect. Note that a complete analysis with real
data will need to take both effects into account.

When a galaxy survey is performed it is vital to distinguish between 
the correlations that we measure in the galaxy distribution and how 
that distribution relates to the distribution of dark matter. 
We assume throughout that the non-linear galaxy power spectrum
is related to the linear matter power spectrum via 
$P_g(k)=b^2P(k) + P_{shot}$. This is motivated by the halo model
\citep[e.g.][]{2000MNRAS.318..203S,2000MNRAS.318.1144P},
a galaxy population will most likely not trace matter in an unbiased way
and given their discreteness and number density there will be an 
extra shot noise power attached to the power spectrum.

Therefore in order to estimate cosmological
parameters from a galaxy catalogue we first choose a fiducial cosmology
which will allow us to change the coordinates from (RA,DEC,z) to spatial
coordinates in redshift space. If, presumably by luck, we choose the
real cosmology that governs our Universe, then the power spectrum that
we will measure as a result of the galaxy autocorrelation function is

\begin{equation}
P_{obs}(k,\mu) = b^2 g^2(z)
\left(1+\frac{f_{k}\mu^2}{b} \right)^2 P(k) +P_{\rm shot}.
\label{eq::pk_true}
\end{equation}

However if we chose an incorrect reference cosmology, then we will observe 
the signal produced from the power spectrum derived from the real
cosmological parameters distorted by the incorrect assumption
we have made. This results in the following expression, for 
what measurements of $P(k)$ we would retrieve

\begin{eqnarray}
P_{obs}(k_{\rm ref},\mu_{\rm ref}) =
\frac{D^2_A(z)_{\rm ref} H(z)}{D^2_A(z) H(z)_{\rm ref}} b_{\rm true}^2
\left(1+\frac{f_{k}\mu^2}{b} \right)^2 
\nonumber
\\
\,\,\,\,\,\,\,\,\,\,g_{\rm true}^2(z) 
P_{\rm true}(k) +P_{\rm shot}.
\label{eq::pk_false}
\end{eqnarray}

\noindent Hence if an incorrect cosmology is assumed there will be an 
inconsistency when it comes to compare the power spectrum expected
given the assumptions we made about the cosmological parameters. With
the power spectrum derived from the data given assumed distances,
we will be able to use this inconsistency to probe cosmological parameters.
In order to do this, with real data, 
we simply need to repeat this process for may 
different cosmological parameters and find which cosmological parameters 
do not produce any inconsistency by comparing predicted models 
and measured models assuming the cosmology we are testing.

In both equations \ref{eq::pk_true} and \ref{eq::pk_false} a relationship
between ($k,\mu$) and ($k_{\rm ref}, \mu_{\rm ref}$) 
depends on cosmology in the 
following way. 

\begin{equation}
\begin{array}{ll}
k^2 = k^2_{\perp}+ k^2_{\parallel} & 
k^2_{\rm ref} = k^2_{\perp,\rm ref}+ k^2_{\parallel,\rm ref} \\
k_{\parallel} = \mu k & k_{\parallel,\rm ref} = \mu_{\rm ref} k_{\rm ref} \\
k_{\perp,\rm ref}D_{A,\rm ref}=k_{\perp}D_A &  
k_{\parallel,\rm ref}H_A=k_{\parallel}H_{A,\rm ref} 
\end{array}
\end{equation}

\noindent which produce the following relations that we use to convert from
a given angle and $k$ value in the sky to another distorted angle and $k$ value
if the incorrect cosmology is chosen

\begin{equation}
\begin{array}{l}
\displaystyle{\mu^2 = \frac{\mu_{\rm ref}^2}{\mu_{\rm ref}^2+
(1-\mu_{\rm ref}^2)
\frac{D_{A,\rm ref}(z)H_{A,\rm ref}(z)}{D_{A}(z)H_{A}(z)}}}\\
\displaystyle{k = k_{\rm ref}\frac{H(z)}{H_{\rm ref}(z)}
\frac{\mu_{\rm ref}}{\mu}}.
\end{array}
\end{equation}

\noindent This is a thin shell approximation and strictly 
speaking should not be used 
in this analysis. However, we argue that the effect on the power spectrum
of using a thin shell 
approximation, if the shells used are thick, is equivalent to
considering a power spectrum convolved with a narrow $k$ 
space window function. We therefore consider,
for forecasting purposes, this is a good enough approximation as shown by 
\citet{2005ApJ...631....1G} for similar survey geometries.

Here we will produce forecasts using Markov Chain Monte Carlo (MCMC) methods to
predict the posterior probability distribution for a given set of cosmological
parameters. In this case, given that no real data are available, we proceed 
in the following way. We chose a fiducial cosmology which we will assume 
to be close to what we think reproduces the real distribution of galaxies.
In order to produce a MCMC that will be able to 
provide us with the error forecasts we need to have a prescription to
compare the log(likelihood) difference between two potential models.
We compare a model with another model which has been distorted
by the incorrect choice of cosmology.
In other words in order to produce a log(likelihood) 
difference between two models we compare
the results from Eqn.\ref{eq::pk_true} with results from Eqn.\ref{eq::pk_false}
having assumed that the errors in the power spectrum will be given by 
Eqn.\ref{eq::error_pk}. 
We calculate the $\chi^{2}$ for simulated galaxy surveys by using a method
involving bins in $\mu$, redshift $z$ and $k$, and including only $k$ modes up to a 
conservative value of $k_{max} = 0.2 ~ h ~ {\rm Mpc}^{-1}$, beyond which the
power spectrum becomes non-linear.

\subsection{Methods for CMB (Planck) datasets}

We will briefly describe how we use MCMC methods to produce forecasts
for future CMB experiments and specifically for the angular power spectrum measurement 
to be achieved by the CMB satellite Planck (http://www.rssd.esa.int/Planck).
Planck, like WMAP before it \citep{2003ApJS..148..175S,2006astro.ph..3449S}, will  produce a CMB map
by using several radio bands to remove foregrounds, especially dust and synchrotron emission
arising from the Milky Way and point sources. The nearly constant background corresponding to a
2.7 K black body will be removed, as will the pure dipole component due to motion with respect to the CMB frame.
This leaves a map of CMB anisotropies $\Theta(\theta,\phi)$ which is decomposed into spherical harmonics

\begin{equation}
\Theta(\theta,\phi)=\sum_{(l,m)}a_{lm}Y_{lm}(\theta,\phi).
\end{equation}

The statistical properties of the coefficients $a_{lm}$ are translated into constraints on 
the underlying cosmological parameters which produce the fluctuations 
and determine their evolution. Theories for cosmic inflation \citep[e.g.][]{1981PhRvD..23..347G} predict that
the primordial density fields are Gaussian in nature, in which case the average $\left<a_{lm}\right>=0$
and it is the variance

\begin{equation}
\begin{array}{c}
\left<|a_{lm}a^*_{l'm'}|\right>=\delta_{lm}\delta_{l'm'}C_l
\end{array}
\end{equation}

\noindent that contains the crucial cosmological information.

The values of $C_l$ can be calculated from a `Boltzmann 
code' such as CAMB \citep{2000ApJ...538..473L} or CMBfast 
\citep{1996ApJ...469..437S}; in this work we use CAMB.

To produce forecasts, we need to have a good idea of the
errors. Even in the absence of any experimental errors,
there is an intrinsic `cosmic variance' because there is a limited number of modes
(i.e.\ a finite number of $a_{lm}$ for a given $l$) 
that can be measured on the sky. 
The total number of $a_{lm}$ is simply 
related to the total number of independent pixels 
available in the sky anisotropy map. The cosmic variance error
can be written as

\begin{equation}
\frac{\Delta C_l}{C_l} =\sqrt{\frac{2}{f_{sky}(2l+1)}},
\end{equation}

\noindent where the factor $2l+1$ is the number of modes $a_{lm}$ 
measurable from the data which is proportional to the fractional
area of sky $f_{\rm sky}$ studied and the factor $2$ reflects that fact that the 
direction of the modes on the sky is unimportant. 

We must also account for experimental sources of error on 
the $a_{lm}$ measurements. If we assume that the sources of variance are independent,
which is obviously fine since one is cosmological and the other experimental,
then the variances should be added in quadrature to get the
total error estimate

\begin{equation}
\Delta C_l =\sqrt{\frac{2}{f_{sky}(2l+1)}} (C_l +N_l^2)
\end{equation}

\noindent where $N_l$ will be defined as a function of the experiment.

The experimental error will be determined by the sensitivity $\sigma_c$
and the angular resolution (beam size) $\theta_b$.
Therefore, if we have a certain number of frequency channels observing the
CMB for a given experiment, the noise will be given by the following expression
\citep{1999ApJ...518....2E}

\begin{equation}
\frac{1}{N_l^2} = \sum_{channels} \frac{1}{(\sigma_c \theta_b)^2} 
{\rm exp}\left(-\frac{l(l+1)\theta_b}{8 \rm ln 2}\right),
\end{equation}

\noindent assuming that
the beam smears the measurement on a given $l$ scale.

The sensitivity of the experiment will depend on the 
experimental setup through the equations

\begin{equation}
\sigma_c=\frac{NET}{\sqrt{n_{det}t}}\frac{\theta_{sky}}{\theta_{b}},
\end{equation}

\noindent where $NET$ is the noise effective temperature in one second 
for each 
of $n_{det}$ detectors, $t$ is the integration time and $\theta_{sky}$
is the solid angle of the survey .

The probability of a value of $a_{lm}$ given a $C_l$ value can now be written as

\begin{equation}
p(a_{lm}|C_l)=\frac{1}{\sqrt{2\pi C_l}}
{\rm exp}\, \left(-\frac{a^2_{lm}}{2C_l}\right),
\end{equation}

\noindent so the probability that a given realisation of the sky
is produced, given a set of cosmological parameters, is simply 
the product of all the $a_{lm}$ probabilities which, in the 
case where the values of $a_{lm}$ are not correlated, gives

\begin{equation}
p(sky|C_l)=\prod_{l,m}\frac{1}{\sqrt{2\pi C_l}}
{\rm exp}\, \left(-\frac{a^2_{lm}}{2C_l}\right).
\end{equation}

We can then use Bayes theorem to find that \citep{2002PhRvD..66b3528B}

\begin{eqnarray}
\begin{array}{ll}
\displaystyle{{\rm log} \left( \frac{p_A}{p_B}\right)} & = 
\displaystyle{{\rm log} \left(\frac{p(a_{lm}|A)}{p(a_{lm}|B)}\right)}\\
\,  & = \displaystyle{\frac{f_{sky}}{2}\sum_l (2l+1) 
\left( 1- \frac{C_{l,A}+N_l^2}{C_{l,B}+N_l^2}\right.} + \\
\,  & \,\,\,\,\,\,\,\,\,\,\,\,\,\,\,\,\,\,\,\,\,\,\,\,\,\,\,\,\,\,\,\,\,\,\,\,\,\,\,\,\,\,\,\,\,\,\,\,\,\,\,\,\,\,\,\,\,\,\,\displaystyle{\left.{\rm log}\left(\frac{C_{l,A}+N_l^2}{C_{l,B}+N_l^2}\right)\right)}.
\end{array}
\label{eq:cmb:like}
\end{eqnarray}

However, in the case of future CMB observations like those from Planck, 
we will have access to data which gives information on both
temperature and polarisation anisotropies. 
It is common to separate the polarisation modes in the CMB as E and B
modes which are related to `curl free' and `curl' vector fields. This separation 
is done because most primordial effects generate only E modes, and B modes 
are only generated if there is a gravitational wave background or by
gravitational lensing of E modes. Much additional cosmological information can
be extracted from the E-mode power spectrum as well as the cross correlation
between the E and T (total power) spectra.

\begin{figure*}
\begin{center}
\centerline{
\includegraphics[width=15.5cm,angle=0]{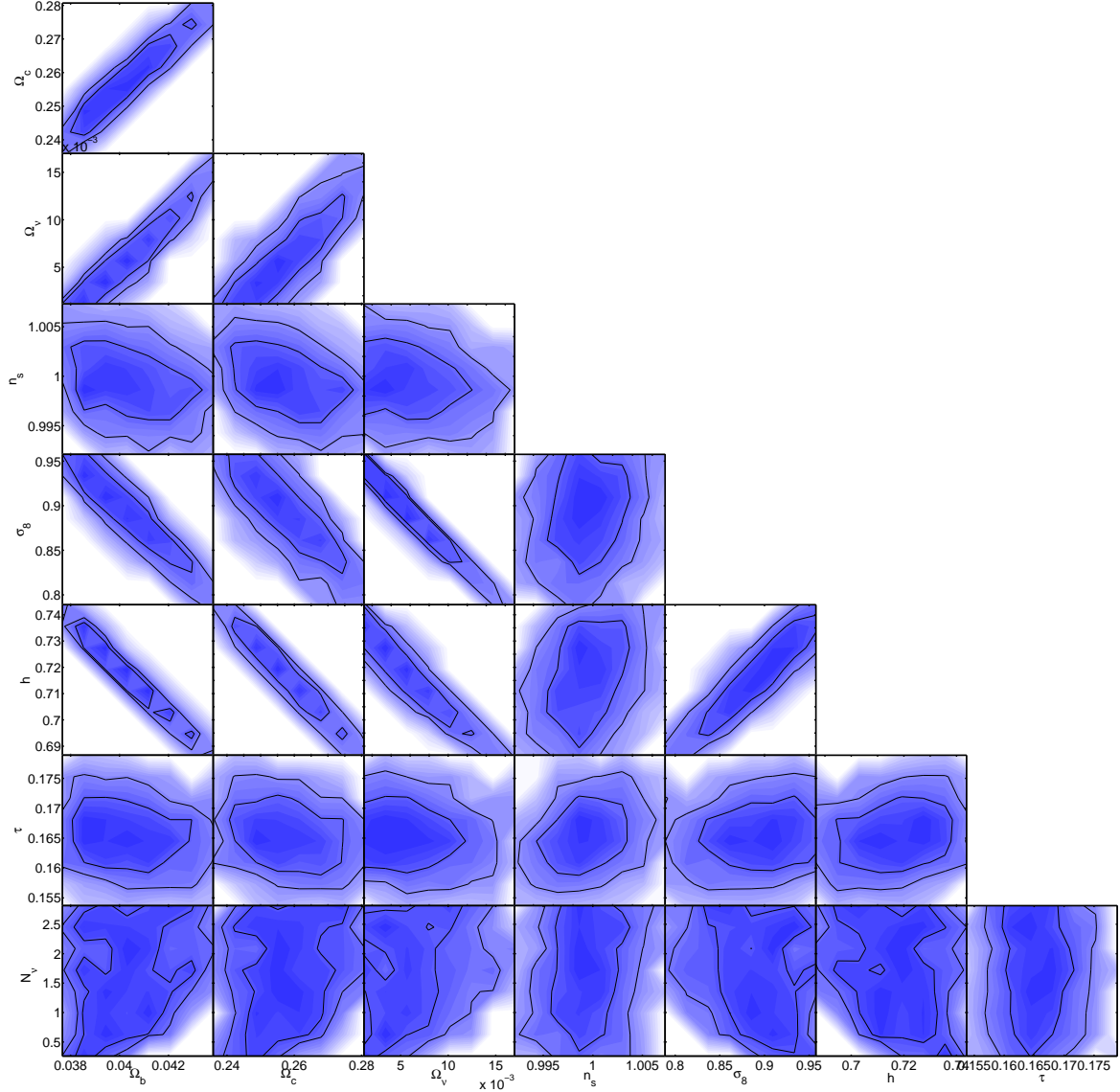}}
\caption[Constraints on quasi-degenerate models from Planck.]{Posterior
probability distribution from an MCMC run determining the 
quality of measurements that the Planck CMB mission 
will have in measuring the absolute neutrino mass scale.
Contours are one sigma and two sigma levels obtained by taking the
68\% and 95\% of the samples which best fit the fiducial model.
It is clear that degeneracies are unlikely to 
allow us to get a clear signal arising from a massive neutrino 
from CMB data alone. Data from Planck
will improve current constraints on the neutrino mass even though a direct 
detection seems unlikely as many models with $\Omega_\nu=0$ are not
rejected at the one sigma level.}
\label{fig::n2_matrix_planck}
\end{center}
\end{figure*} 

Mathematically, Eqn.\ref{eq:cmb:like} is modified in the following way
in the case where the model describes all the power spectra produced from 
random Gaussian fields with a certain degree of cross correlation 
\citep{2002PhRvD..66b3528B}.

\begin{eqnarray}
\begin{array}{ll}
\displaystyle{{\rm log} \left( \frac{p_A}{p_B}\right)} & = 
\displaystyle{{\rm log} \left(\frac{p(a_{lm}|A)}{p(a_{lm}|B)}\right)}\\
\,  & = \displaystyle{\frac{f_{sky}}{2}\sum_l (2l+1) 
\left( Tr(I - M_A M^{-1}_B) + \right.} \\
\,  &  \,\,\,\,\,\,\,\,\,\,\,\,\,\,\,\,\,\,\,\,\,\,\,\,\,\,\,\,\,\,\,\,\,\,\,\,\,\,\,\,\,\,\,\,\,\,\,\,\,\,\,\,\,\displaystyle{\left.{\rm log}\left(Det(M_A M^{-1}_B)\right)\right)}
\end{array}
\label{eq:cmb:like:2}
\end{eqnarray}

\noindent where $I$ is the identity matrix and
the matrices $M_A$ and $M_B$ are given by the values of 
the individual power spectra and their cross correlations at a given 
mode $l$ following

\begin{equation}
M = \left(
\begin{array}{ll}
C^{TT}_l +{N^{TT}_l}^2 & C^{TE}_l\\
C_l^{TE} & C^{EE}_l +{N^{EE}_l}^2
\end{array}
\right).
\end{equation}

We can therefore produce a forecast for any CMB experiment provided we have access
to reliable predictions of the noise level of the experiment for both temperature and 
polarisation. Here, we produce forecasts involving the entire 
likelihood for a Planck-like experiment. 
The Planck specifications used here are given in Table.\ref{tbl::planck1}.

\begin{table}
\begin{center}
\begin{tabular}{|c|c|c|c|c|}
\hline
Frequency & 70 & 100 & 143 & 217 \\
\hline
Beam width $\theta_{b}$ / arcsec & 14.0 & 9.5 & 7.1 & 5.0 \\
NET /($\mu K \sqrt{s}$) & 212(300) & 56(80) & 56(80) & 84(120) \\
Detector number $n_{det}$ & 12 & 8 & 12(8) & 12(8) \\
\hline
\end{tabular}
\caption[CMB (Planck) experimental limits assumed by our forecasts]{Experimental limits used in the forecasts
of future CMB (Planck) data. Values in parenthesis are for polarisation experiments. We 
have assumed one year integration time and a sky coverage of $f_{sky}=0.65$. These data are
taken from the Planck blue book (http://www.rssd.esa.int/Planck)
\label{tbl::planck1}}
\end{center}
\end{table}

We plot in Fig.\ref{fig::n2_matrix_planck} the results we find for an MCMC 
run assuming Planck data alone. 
As we can see the main degeneracies will remain in Planck data, noting 
the strong negative correlation of $\Omega_\nu$ and $\sigma_8$ 
which can be explained by less growth with a cosmology with neutrinos. Specially models in which $\Omega_\nu = 0.0$ cannot be ruled out if values of $m_\nu$ turn out to be near the bottom range allowed by particle physics.
The upper limits derived from Planck data alone 
are better than current estimates for the neutrino mass:
marginalising over other parameters they lie around $\Delta m_\nu \sim 0.65$ eV ($1 ~ \sigma$).
This result is roughly in accordance with other forecasts 
\citep{1999ApJ...514L..65H}, where the authors find a lower value of
$\Delta m_\nu \sim 0.25$ eV ($1 ~ \sigma$) using a Fisher matrix analysis and broadly
similar assumptions.  
In \citet{1999ApJ...514L..65H} the assumptions of noise of the Planck experiment were slightly 
different than the ones assumed here, and their Fisher analysis does not
account of any non-Gaussian distribution of the parameters. \citet{2005PhRvD..71d3001I}
claim that if neutrinos have a mass scale smaller than $\sim 1 ~ \rm eV$ then there
is very little useful constraining information in CMB datasets.

We point out here that we have made an important approximation in our analysis.
We have assumed that we are in the limit where we have such a small mass for the 
lowest mass eigenstate, that we can set the masses of the lowest mass eigenstates to
zero (the two lowest for a normal hierarchy, and the lowest only for an inverted hierarchy;
see Fig.~\ref{fig:neutrino_hyer}). This approximation is valid if the masses
are small enough that we are not in the quasi-degenerate scenario. 
Given that all the scenarios we consider here assume a 
mass below 0.25 eV, this is a reasonable assumption. We have not considered here
the possibility that different neutrino masses may be measurable by cosmological data
as this is beyond the capabilities of the surveys we consider.

\subsection{Results from LSS and effects of priors.}

It was pointed out in \citet{2003JCAP...04..004E} that the role of
priors is vital in retrieving information
about the neutrino mass from cosmological surveys.
This is because of the large set of parameters that determines the
behaviour of cosmological perturbations and the complicated relation between them.
There is for instance a tight correlation between the matter density and the 
neutrino parameter $\Omega_\nu$. This occurs because over most of 
the evolution of the Universe, neutrinos behave
just as cold dark matter, 
hence it is necessary to know the cold dark matter density
well to determine effectively by how much it effectively changes
as neutrinos become non-relativistic.

Given that one of the major effects of a high $\Omega_\nu$ 
is a strong damping of the power 
at small scales due to free streaming, it is strongly 
correlated with the value of $n_s$. 
It is very important to measure $n_s$
in order to deduce a neutrino mass. 
It is possible that the discrepancies between
current results that do agree broadly but not in the detail, 
comes down to the priors used, for instance \citep{2002PhRvL..89f1301E} 
did not assume a varying scalar index in their analysis. They chose to give 
results for different choices of $n_s=0.9,1.0,1.1$. One may 
use theoretical priors to consider only values close to $n_s=1$ but 
in doing so one must be wary of the large degeneracy between $n_s$ and $\Omega_\nu$. 

Most current results rely on priors in one way or another. We argue that in order to 
obtain a clean measurement of the neutrino mass it will be important to use cosmological data
in a way that we are sure that our result in not too prior dependent. We 
will argue in this paper that a future CMB data set, 
such as one from Planck, combined with a 
galaxy survey, such as one from the SKA, can provide the data 
necessary for a robust analysis of neutrinos via cosmology.

\section{Measuring neutrino masses with the SKA}

\begin{figure}
\begin{center}
\centerline{
\includegraphics[width=9.5cm,angle=0]{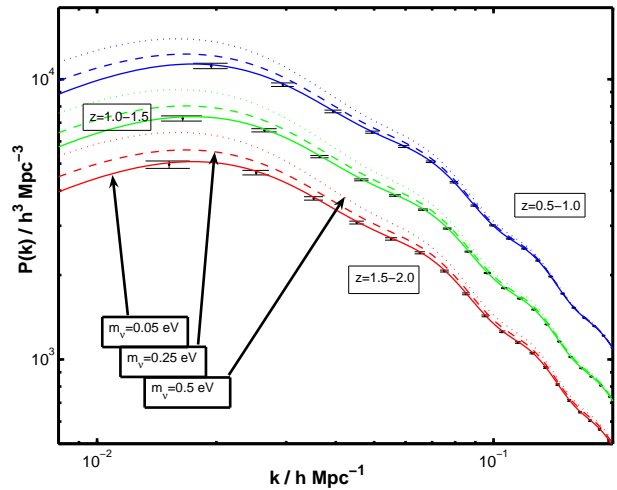}}
\caption[Neutrino signature on SKA data.]{Illustration of the sensitivity
a future LSS (SKA) survey will need to measure and constrain the absolute neutrino mass scale $m_{\nu}$.
The error bars illustrate the accuracy of the LSS measurement assuming the validity of Eqn.\ref{eq::error_pk}, with $P(k)$
taken from a fiducial cosmological model which neglects 
massive neutrinos \{$\Omega_b$, $\Omega_c$, $w$, $h$, $n_s$, $\sigma_8$\}=
\{0.04, 0.26, -1, 0.72, 1.0, 0.9\}. 
Note that the plotted error bars should be totally uncorrelated because of 
the large area coverage and redshift depth, and hence broad real-space window function, of the SKA survey.
The solid, dashed and dotted lines correspond to the addition to the fiducial model of 
neutrinos of mass 0.05 eV, 0.25 eV and 0.5 eV (taking $N_\nu = 3$), all of which are still
possible values given the constraints from current data sets. The different colours represent 
independent SKA measurements in three redshift slices: $0.5 \leq z < 1.0$ (blue), $1.0 \leq z < 1.5$ (green) and $1.5 \leq z < 2$ (red). 
\label{fig::neutrino_ska}}
\end{center}
\end{figure}

\begin{figure*}
\begin{center}
\centerline{
\includegraphics[width=15.5cm,angle=0]{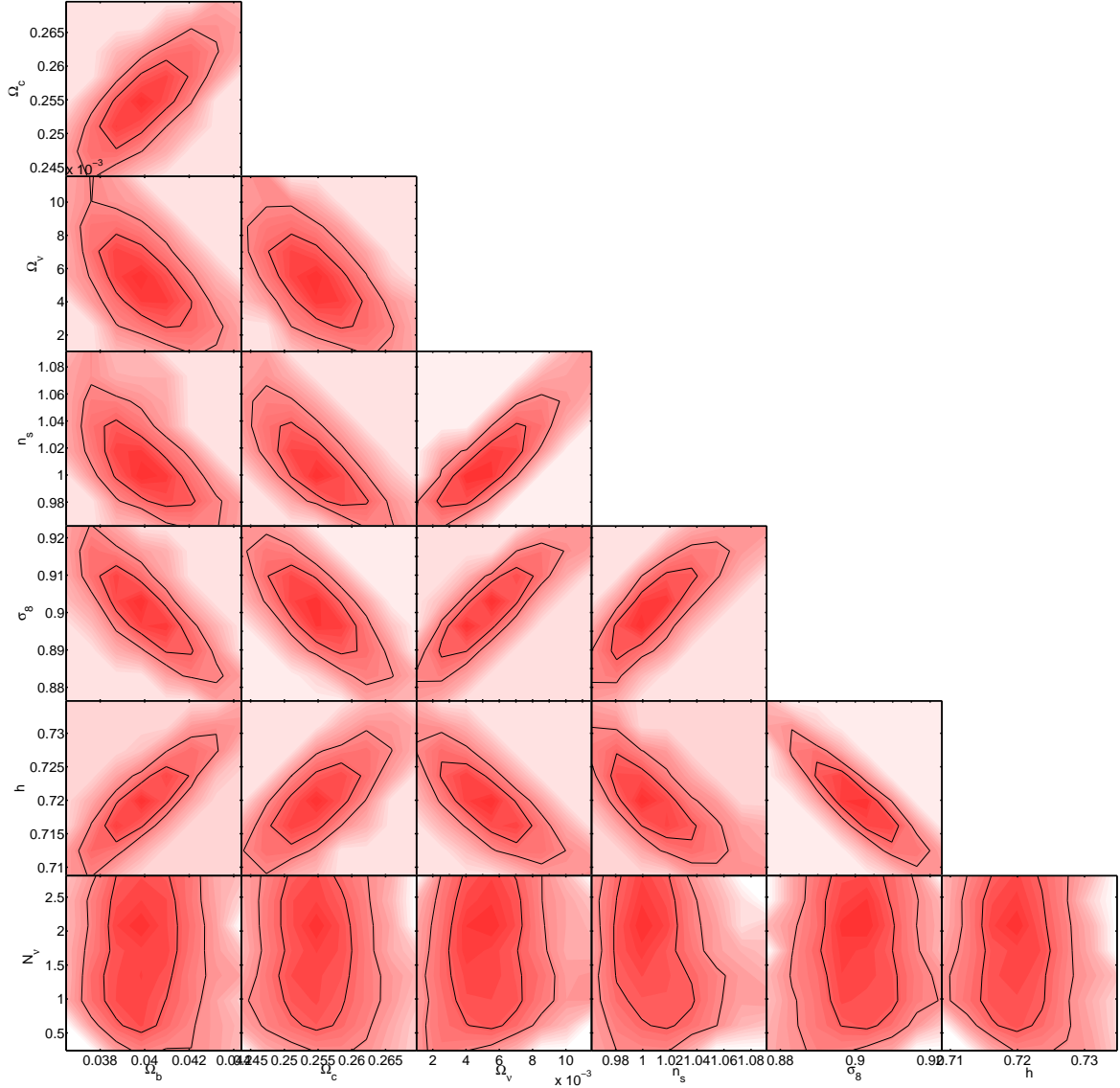}}
\caption[Constraints on quasi-degenerate models from SKA.]{Analysis 
of the accuracy that a future SKA survey will have 
in constraining a neutrino mass scale. Contours are one sigma and two sigma levels obtained by taking the
68\% and 95\% of the samples which best fit the fiducial model.
The SKA alone 
will be able to, by accurately measuring the shape of the
power spectrum in several different redshift bins, measure a signal from the neutrino energy density
at more than the three sigma level by marginalising over other parameters and without 
any further priors needed by current data sets. }
\label{fig::n2_matrix_ska}
\end{center}
\end{figure*} 

\subsection{A back-of-the-envelope calculation}
\label{sec:bec}
As we have outlined, a measurement of the shape of the power spectrum at 
large scales as given by Eqn.\ref{eq::nu_pk} 
would be able to constrain $m_{\nu}$ for 
the neutrino sector. Here we attempt to estimate the extent to which
we would be able to measure this effect with a future large-scale-structure (LSS) survey.
In this section we outline a simple back-of-the-envelope calculation to estimate
the cosmic volume required, and then in Sec.\ref{mcmc::quasi} we undertake a more 
detailed calculation that takes into account the full effect that 
neutrinos have both on LSS and CMB data.

However, data produced by redshift surveys on large scales is often
cosmic-variance limited. In this case, the fractional error with which we can 
measure $P(k)$ is proportional to the number of $k$ modes present in the survey volume, 
and hence the accuracy is inversely proportional to the square root of the 
cosmic volume (Eqn.\ref{eq::error_pk}).

If we ignore for now the role of priors that, in 
reality, complicate the analysis (Sec.~\ref{mcmc::quasi}), 
the upper limits from
surveys such as the 2dF Galaxy Redshift Survey (2dFGRS) and the SLOAN Digital Sky Survey (SDSS)
are around $m_\nu \sim 1{\rm eV}$ \citep{tegmark-2006-}. Hence, a survey with $\sim 400$ times the cosmic
volume would have errors on large scales a factor of $\sim 20$-times smaller and therefore would be able to 
probe $m_{\nu} \sim 0.05$ eV at the bottom end of the range allowed by particle physics,
provided the relation given in Eqn.\ref{eq::nu_pk} holds. Such an experiment would, 
by detecting the signal from neutrinos imprinted on the large-scale structure, be able to
probe the entire neutrino sector allowed by current particle physics experiments

Furthermore, if we do have such a large volume to search for features imprinted on 
the galaxy power spectrum, other cosmological parameters will also be accurately determined. 
The problems with priors and parameter degeneracies that plague current analyses should be much reduced. 
Current redshift surveys also suffer from the problem of correlated errors on large scales because window functions 
of these surveys are necessarily narrow in real space, and hence broad in $k$ space. An `all sky' survey reaching to
very high redshifts would be the optimal way to probe such signals, but we will consider here an, as yet
hypothetical, $20,000 ~ \rm deg^{2}$ LSS survey reaching to redshift $z \sim 2$. Such a survey would provide the
huge increase in cosmic volume needed to comprehensively probe neutrino properties. For specific calculations we
will use simulated LSS datasets from surveys with the SKA, although clearly our conclusions will be valid
for any LSS survey with similar reach in sky area, redshift depth and galaxy number density. Further details 
concerning future redshift surveys with the SKA can be found in the following references: details
concerning galaxy number densities in \citet{2005MNRAS.360...27A}; details concerning $P(k)$ measurement
in \citet{ABR}; a comparison with other future LSS datasets in \citet{RA}.

\subsection{The cosmological imprint of neutrinos in a future LSS data set}

We illustrate in Fig.\ref{fig::neutrino_ska} the promise of a future LSS (SKA) survey for detecting the
imprint of neutrinos on the power spectrum $P(k)$. We note that we have chosen bins in redshift space which correspond to intervals
in comoving space larger than the $\lambda$ corresponding to $k=0.01 h^{-1} {\rm Mpc}$, hence the theoretical error bars plotted are uncorrelated. Fig.\ref{fig::neutrino_ska} is the result of a simple analysis in which all 
other cosmological parameters are fixed at fiducial values and $m_\nu$ is varied through the 
neutrino density parameter $\Omega_\nu$ (using Eqn.\ref{eq::omega_nu_nb} in a quasi-degenerate scenario with $N_\nu = 3$). We can see 
clearly that this 
agrees well with the back-of-the-envelope calculation presented in Sec.\ref{sec:bec}, with 
even models with $m_{\nu} \sim 0.05 ~ \rm eV$ starting to show significant deviations from the
fiducial model which ignores neutrinos.

\begin{figure}
\begin{center}
\centerline{
\includegraphics[width=9.5cm,angle=0]{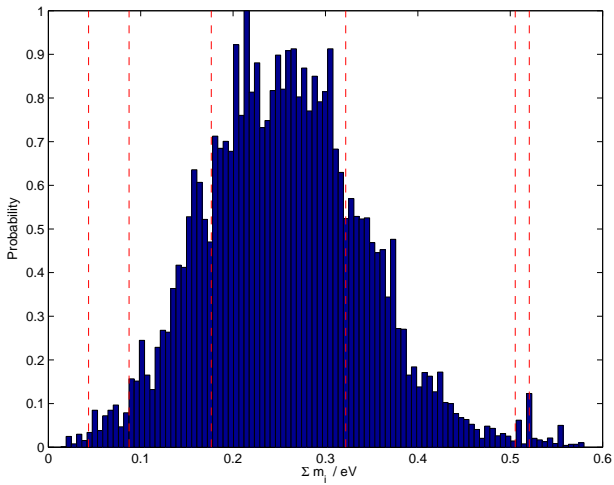}}
\caption[Confidence levels obtained assuming $m_\nu = 0.25$ eV and SKA data.]{Analysis 
of the accuracy that a future SKA survey will have 
in constraining an absolute neutrino mass $m_\nu \sim 0.25$ eV. 
The plot shows the forecast of the
unnormalised probability one expects to have for the sum of the neutrino 
masses. Here the fiducial value has been chosen to be $\Sigma m_i = 0.25$ eV.
The vertical red dashed lines correspond to one,
two and three sigma confidence levels.
By accurately measuring the shape of the
power spectrum in three different redshift bins,
the SKA alone will be able to,  
measure a signal from the neutrino mass
at more than three sigma level by marginalising over other parameters and without 
any further priors as are needed for current data sets.}
\label{fig::m_nu2_ska}
\end{center}
\end{figure} 

Another key point illustrated by Fig.\ref{fig::neutrino_ska} is the importance of undertaking
a LSS survey in independent redshift shells. A common criticism of attempts to constrain $m_{\nu}$ from
LSS measurements is that the feature being looked for maybe being masked, or even mimicked, by
features introduced by the use of galaxies as an indirect probe of the underlying fluctuations. 
A LSS survey which can probe several 
redshift shells can give a very good handle on any such systematic errors present in a galaxy redshift survey. 
One specific worry that has to be addressed is whether any biasing effect from the galaxy power 
spectrum would be able to produce any scale-dependent modulation mimicking the damping tail of massive neutrinos \citep{2000MNRAS.318..203S}. First, any worry on this specific issue is attenuated from the theoretical 
expectation from the halo model (\citet{2000MNRAS.318..203S,2000MNRAS.318.1144P}; and see \citet{ABR}) that the galaxy bias 
is, at least on large scales, reliably modelled as a constant multiplicative factor plus a constant additive
shot noise. It is only at small scales that this assumption should break down. Hence, the neutrino mass 
estimate should be robust against this kind of systematic. Second, as illustrated in 
Fig.\ref{fig::neutrino_ska}, the huge number of galaxies of an SKA-like LSS survey would allow 
us to probe the galaxy power spectrum with several galaxy types in each redshift
shell, and therefore produce an independent power spectrum for each galaxy type. If we are 
concerned about any systematic problem due to galaxy bias, we should be able to compare the power spectrum for different galaxy types in each shell.
All shells should have the same signal arising from neutrino physics, whereas each
power spectrum would be different if there is an effect mimicking the effects of neutrinos 
because each is made up of samples of galaxies which have different bias and clustering 
properties. \citet{2006astro.ph.11178C} have shown that it is important to correctly model 
any scale dependent bias which is luminosity dependent. With a future LSS survey it will be important 
to have enough galaxies to correctly model this effect. With an SKA survey the number of galaxies will 
be large enough 
so that we are able to model the scale dependence of the bias as a function of galaxy properties which 
we can retrieve from the data, i.e. as a function of hydrogen mass and circular velocity which should 
correlate to the dark matter mass of the galaxy. There is therefore a better prospect to model a scale 
dependent bias than may be possible with optical surveys. 

\begin{figure}
\begin{center}
\includegraphics[width=7.7cm,angle=0]{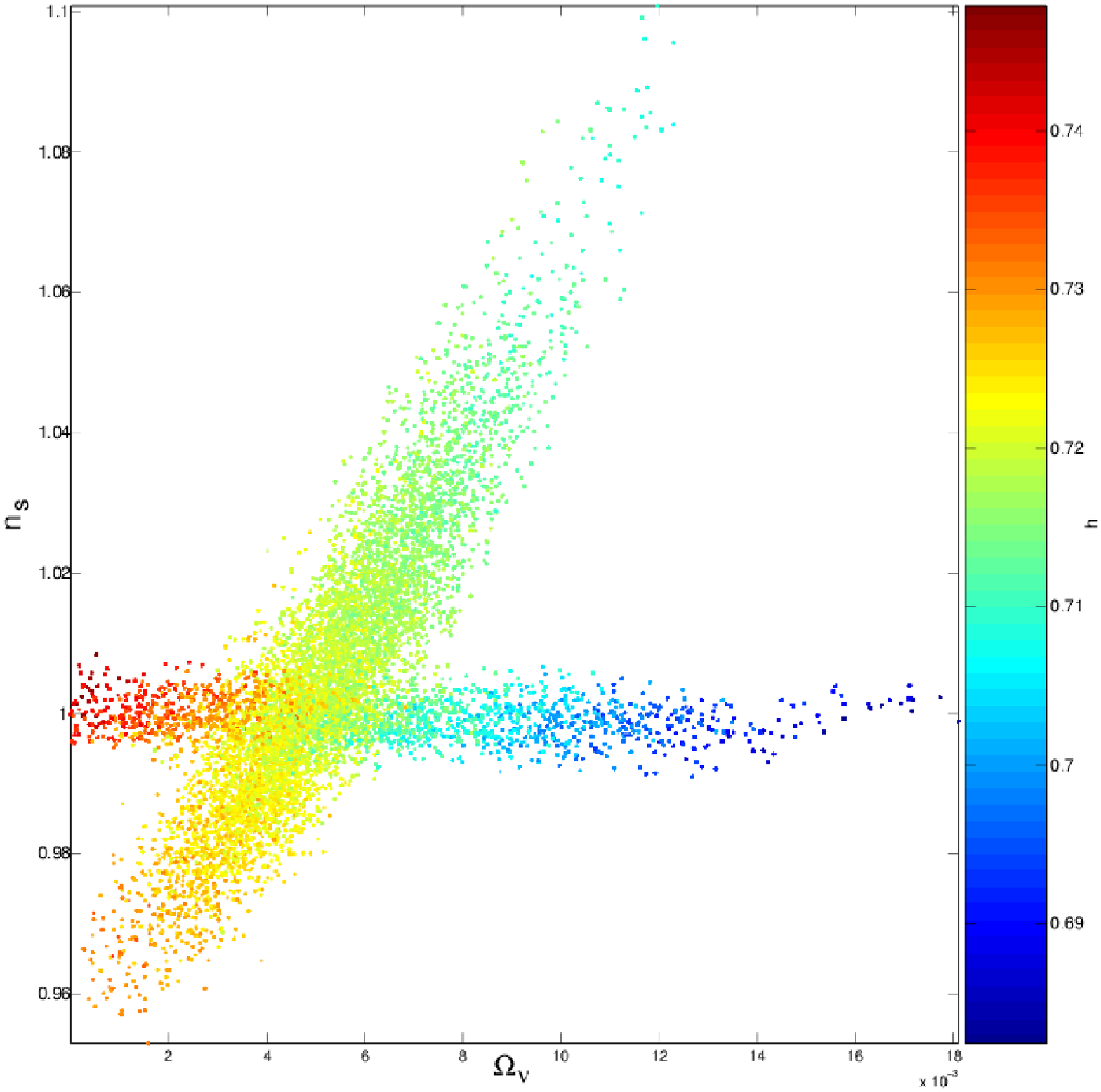}
\centerline{
\includegraphics[width=7.7cm,angle=0]{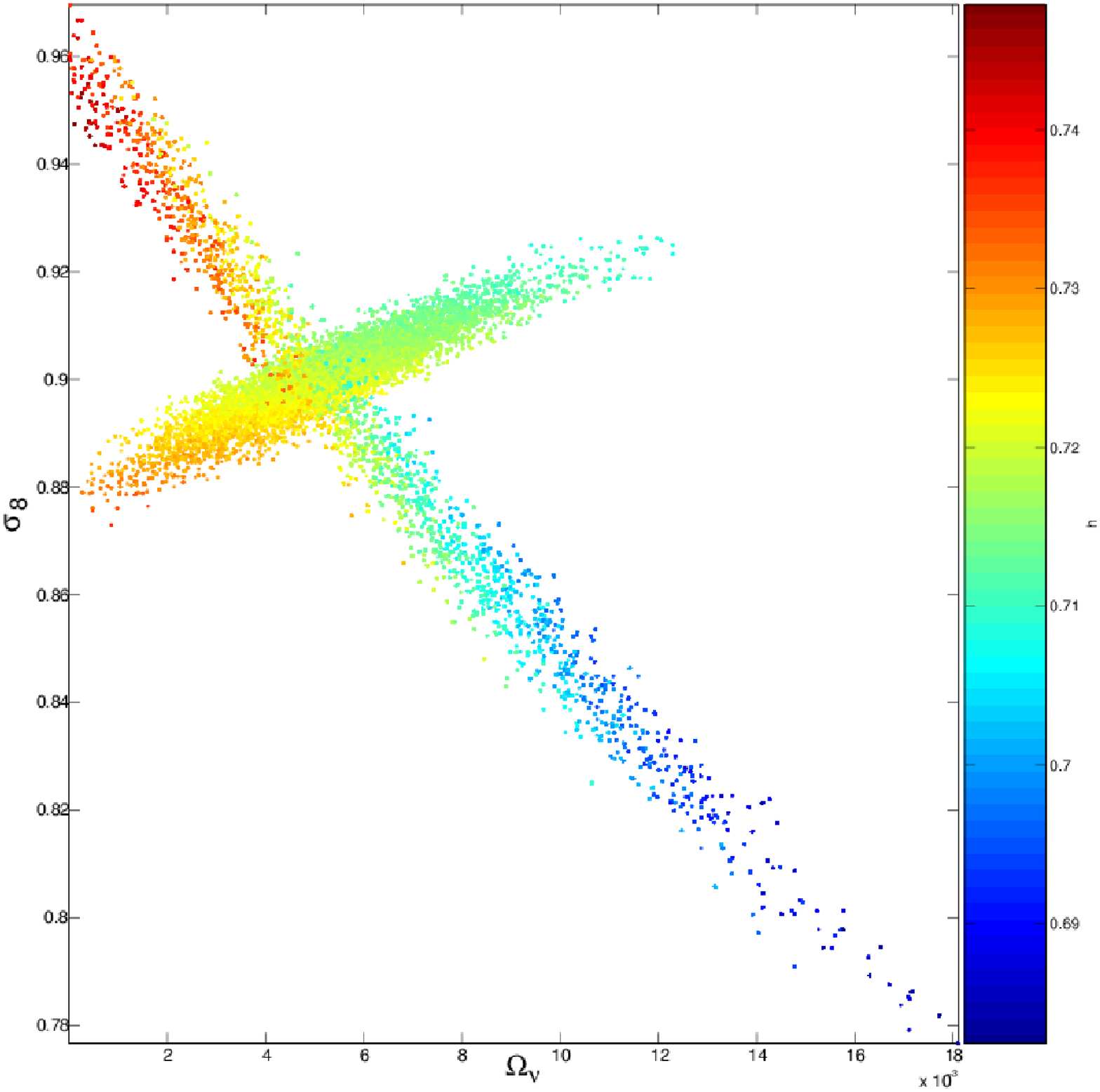}
}
\caption[Cosmic complementarity between LSS and CMB results.]{We 
illustrate the 
main degeneracies present in LSS (SKA) and CMB (Planck) data and how the combination of both 
approaches can improve the 
errors on $\Omega_\nu$. We plot MCMC samples from both experiments, separately
simulated, with different parameters along the x and y axis. The top panels illustrate that
the CMB will measure $n_s$ accurately whereas there is a degeneracy
between $\Omega_\nu$ and $n_s$ in LSS data. The bottom panel shows that there
is an anti-correlation between the parameters $\sigma_8$ and $\Omega_\nu$
for CMB data whereas there is a positive correlation in the case of LSS data;
this is due to different effects: for the CMB, the 
$\Omega_\nu$ - $\sigma_8$ degeneracy is probing the
amount of growth from recombination to the present day; for the LSS, the 
$\Omega_\nu$ - $\sigma_8$ degeneracy is due to the 
shape of the power spectrum which is determined by $\Omega_\nu$ 
and which is central to the determination of $\sigma_8$.}
\label{fig::degeneracies1_ska_cmb}
\end{center}
\end{figure} 

However, there is another potential problem that Fig.\ref{fig::neutrino_ska} fails to address.
We still need to test whether constraints on neutrino parameters are robust to
problems induced by degeneracies between cosmological and neutrino parameters. Such degeneracies could
make an LSS experiment more heavily reliant on priors than on the data itself, and hence
yield misleading forecasts.

\subsection{MCMC methods probing a quasi-degenerate scenario with the SKA.}
\label{mcmc::quasi}

In order to probe the cosmological parameter space for these models we use a Markov Chain Monte Carlo (MCMC)
methods in order to obtain a prediction of the posterior probability
for a given model and
a given future survey. We use a $\Lambda CDM$ model with neutrinos as a model cosmology
and use the following 13 parameters in our MCMC chains: \{$\Omega_b$, $\Omega_c$, $\Omega_\nu$, $h$, $n_s$, $\sigma_8$, $N_{\nu}$, $b_1$, $p_1$, $b_2$, $p_2$, $b_3$,$p_3$\}.
The first seven parameters are the 
usual cosmic parameters and the last six parameters 
are the multiplicative bias and an additive
shot noise power $P_{shot}$ arising from the halo model, in each one of the three redshift bins,
that we marginalise over.

We will focus on two MCMC-based studies: one which assumes the absolute 
mass scale for the neutrinos $m_\nu \sim 0.25\, {\rm eV}$ chosen to be the 
lowest mass which an SKA survey can, on its own, measure sufficiently
to discriminate between a normal and inverted hierarchy; another which 
combines Planck and SKA data on the assumption that the mass scale is 
$m_\nu = 0.05\, {\rm eV}$, the lowest allowed by particle physics.

\subsubsection{SKA-only study at $\Sigma m_i \sim 0.25$ eV.}
\label{ssec:ska_nu}

We first choose to examine whether a survey with the SKA will be able to probe $m_\nu$
as low as $\sim 0.25\, {\rm eV}$ and be able to completely rule out models which have a 
quasi-degenerate neutrino mass spectrum. We therefore choose a fiducial model 
which has only two massive neutrinos with the following cosmological
parameters 
\{$\Omega_b$, $\Omega_c$, $\Omega_\nu$, $h$, $n_s$, $\sigma_8$, 
$N_{\nu}$\} =
\{0.04,0.255,0.005,0.72,1.0,0.9,2.0\}. 
We impose an uniform 
top-hat prior on $0 < N_\nu < 3$ assuming that the standard model
for neutrinos imposes three neutrino eigenstates.
The fiducial bias for all the redshift bins has been 
conservatively set to one but we note that if a larger bias is found, which is 
likely to be the case, 
especially in higher redshift bins the significance of 
the measurements of neutrinos will improve. We assume a fiducial shot noise
value equal to one over the number density of galaxies.

\begin{figure*}
\begin{center}
\centerline{
\includegraphics[width=15.5cm,angle=0]{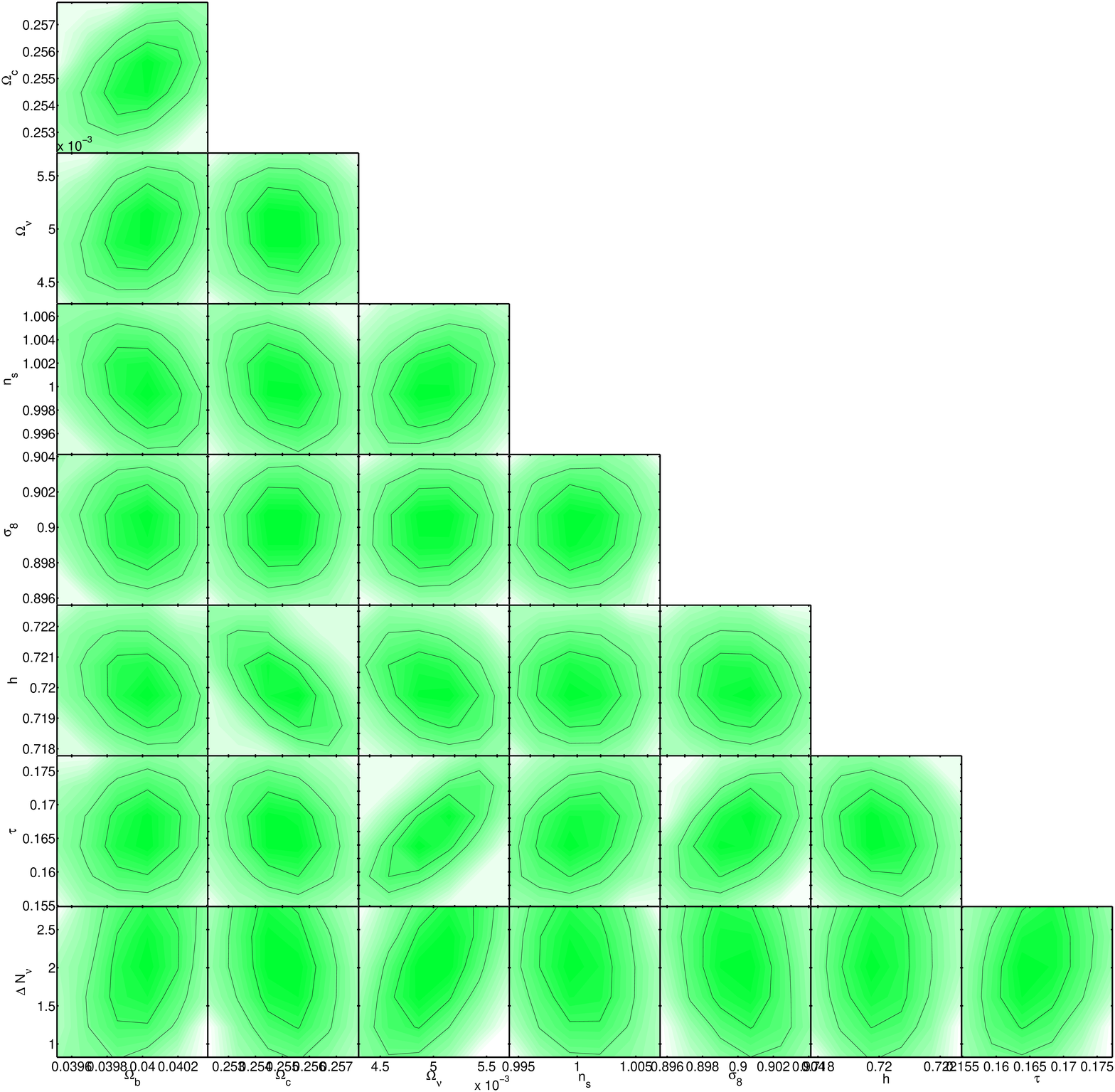}}
\caption[Constraints on quasi-degenerate models from SKA.]{Similar figure to
Fig.\ref{fig::n2_matrix_ska} for a LSS (SKA) plus CMB (Planck) experiment. 
Contours are one sigma and two sigma levels obtained by taking the
68\% and 95\% of the samples which best fit the fiducial model.
The combination of these two experiments will be able to constrain several
cosmological parameters to the 1 per cent level. Most notably, the  
energy density in neutrinos will be constrained to $\sim 0.015$ eV and the 
number of massive neutrino species to $\sim 0.5$.}
\label{fig::n2_matrix_both}
\end{center}
\end{figure*} 

We plot in Fig.\ref{fig::n2_matrix_ska} and Fig.\ref{fig::n2_matrix_both} 
the MCMC 
samples for an analysis following the methods outlined. 
We note that with this future SKA survey alone, there should 
be a $\gtsimeq 3 \sigma$ 
detection of the absolute mass scale of the neutrino. 
The measurement of the power spectrum is accurate enough
that other parameters are well constrained so stringent
priors are therefore unnecessary for a detection of the neutrino signature
to be made. We argue that 
this is of extreme importance for a result to be considered robust not only
by the astrophysical community but also by the particle physics community.
For such a sample we can estimate the direct forecasted posterior on the sum 
of the neutrino masses. We plot this in Fig.\ref{fig::m_nu2_ska} 
from which the neutrino absolute 
mass scale can be detected at more than the three sigma level.

\subsubsection{SKA plus CMB study at $\sum m_i \sim 0.25\, {\rm eV}$.}
\label{ssec::ska_lc_m_nu_025}

It is important to understand and predict how 
other data will affect the predictions of Sec.\ref{ssec:ska_nu}.
Even though we have argued that 
restrictive priors may be unhelpful when it comes to 
interpreting how much information is encoded in a given data set,
it is certain that additional data sets can provide a huge gain in 
sensitivity for a given cosmological
experiment as has been pointed out by many authors 
\citep[e.g.][]{1998PhRvL..80.5255H}.

If we ignore the problems of parameter degeneracy 
a LSS survey which can get redshifts for sources over $20,000\, {\rm deg}^2$ out to
a redshift of two would be able to reach $m_\nu \sim$ 0.05 eV (Fig.\ref{fig::t_k_neutrino}).
However the problem of parameter degeneracy plagues any interpretation of these data.
For instance if we are assessing models which have small values of 
$\Omega_\nu$ then the suppression of the
power at small scales is very small and can be mimicked by a change in $n_s$ 
in the primordial power spectrum. Hence a LSS survey would be unable to
distinguish between a 
signal left over from an inflationary phase of the Universe
and a signal due to neutrinos.

\begin{figure}
\begin{center}
\centerline{
\includegraphics[width=8.5cm,angle=0]{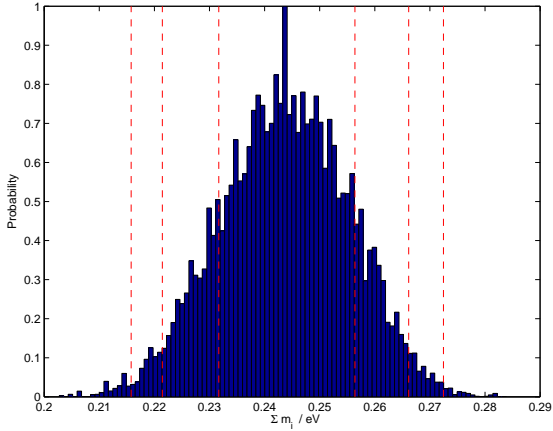}}
\caption[Confidence levels obtained for the neutrino mass assuming $m_\nu = 0.25$ eV,
SKA and Planck data.]{Confidence levels obtained for the neutrino mass assuming $m_\nu = 0.25$ eV for
LSS (SKA) and CMB (Planck) data. A comparison between this figure and 
Fig.\ref{fig::m_nu2_ska} shows how the accuracy of the measurement is improved by a factor of five
when both experiments are considered. Vertical red dashed lines correspond to one,
two and three sigma confidence levels.}
\label{fig::m_nu2_both}
\end{center}
\end{figure}

In Fig.\ref{fig::n2_matrix_both} we see the full posterior probability distributions
for the combination of LSS (SKA) and CMB (Planck) experiments. 
We illustrate in Fig.\ref{fig::degeneracies1_ska_cmb} the $\Omega_\nu$ - $n_s$ degeneracy 
present in a future SKA survey on its own. This degeneracy is made worst 
because we have 
included $k$ modes up to a 
conservative value of $k_{max}$ of 0.2 $h ~ {\rm Mpc}^{-1}$. 
The power spectrum becomes non-linear at larger scales (smaller k) than 
$k_{max}$ and if we could find a satisfactory way 
of modelling these non-linearities
we would be able to use these data to help break this degeneracy.
However Planck will be able to probe high $k$ values by measuring
high $l$ values in the CMB power spectrum. As gravitational growth is much 
less advanced at $z = 1000$, non-linearities are not an issue and CMB 
data are therefore also ideal to constrain $n_s$. As we
can see from Fig.\ref{fig::degeneracies1_ska_cmb}, Planck will measure $n_s$
very well independently of most other parameters. Hence a combination of these
two data sets will improve the error bars on the neutrino mass by a 
factor of five assuming
a fiducial model with $m_\nu = 0.25$ eV. We plot the improvement 
on the estimates in 
Fig.\ref{fig::m_nu2_both}.

Another degeneracy that plagues estimates of the neutrino mass via 
cosmological methods is the 
value of the initial fluctuations which can be parameterised as the value it has at the 
CMB via $A_s$ or the current strength of the anisotropies parametrised by $\sigma_8$. 
Given that one of the effects of a massive neutrino in cosmology is the change 
in the growth
factor for a lengthy period of the history of the Universe, a direct comparison between the 
scale of anisotropies on the CMB and more locally with redshift surveys can also improve greatly
the estimates of the neutrino mass. 

We plot in Fig.\ref{fig::degeneracies1_ska_cmb}
the $\sigma_8$ - $\Omega_\nu$ degeneracy for CMB and LSS surveys. 
Whereas for the CMB, 
growth is responsible for the large degeneracy between $\Omega_\nu$ and $\sigma_8$, 
for the a LSS survey is is the
shape of the power spectrum which is mainly 
responsible for the observed degeneracy, i.e. $\sigma_8$ fixes a weighted 
integral under $P(k)$, whereas $\Omega_\nu$ changes the 
overall shape of $P(k)$. By combining the two
estimates a huge improvement can be achieved. 

However a poor knowledge of the bias parameter would not allow us to combine these data sets
in a consistent way and the improvement would be much smaller. We have assumed that the bias
of high redshift galaxies can be measured and marginalised via measurements
of modes with different sky orientation, through the influence of 
redshift space distortions (see Eqn.\ref{eq::pk_true}). 
In our MCMC chains the bias is constrained to a few per cent level
and marginalised over. For a more realistic simulation we could include other effects 
such as small scale
redshift space distortions. Current spectroscopic
surveys with much smaller samples are able to measure the bias to about the 
several per cent level by
measuring bias using higher order statistics such as the bispectrum \citet{2002MNRAS.335..432V},
so we argue that the assumptions we have made here are not over-optimistic.

However we point out that it is possible that the data at high redshift
will allow us to probe linear scales up to a much larger $k$ values in which case the constraints 
coming from LSS alone would be greatly improved. Also a better understanding of non-linearities 
coming from the halo model as well as N-body simulations \citep{2005Natur.435..629S} are very promising, 
and there are realistic hopes that 
we could use data from the mildly non-linear 
regime without introducing large systematic errors and hence improve 
the constraints shown here.

\subsubsection{Directly probing neutrino hierarchies with cosmological data with $\sum m_i \sim 0.25 ~ \rm eV$.}
\label{sec::direct_hi}

In our analysis of mock future data we have assumed that the number of 
massive neutrinos $N_\nu$ is constant.
If we assume the standard scenario for neutrinos 
this can arise in a hierarchical case when the
hierarchy is either an inverse hierarchy or a normal 
hierarchy (Fig.\ref{fig:neutrino_hyer}).
If the hierarchy is quasi degenerate, then the number of massive neutrinos 
will be characterised by $N_\nu = 3$ and this represent a density 
of neutrinos equal to 422 million particles per ${\rm m^3}$.

\begin{figure}
\begin{center}
\centerline{
\includegraphics[width=9.5cm,angle=0]{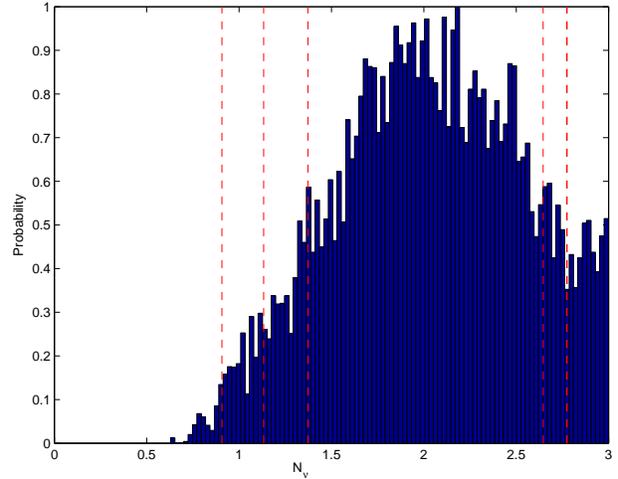}}
\caption[Confidence limits on neutrino hierarchies with SKA + Planck data.]{Confidence limits on neutrino hierarchies with LSS (SKA) and CMB (Planck) data. 
We assess whether a future survey can measure the number density
of massive neutrinos and hence have a direct measurement of the hierarchy of the neutrino 
mass scales assuming a total
energy density in relativistic particles to be $\Omega_\nu = 0.005$ which corresponds roughly 
to a neutrino mass of 0.25 eV. 
The vertical red dashed lines correspond to one,
two and three sigma confidence levels.
We can see that the parameter $N_\nu$ is a difficult parameter to measure and 
even with exquisite data coming from future galaxy surveys and future CMB experiments
it will be hard to constrain this parameter to better than a $\delta N_\nu > 1$. 
For this specific scenario the value $N_\nu = 1$ can be rejected at 2.7 $\sigma$.}
\label{fig::n2_both_ska}
\end{center}
\end{figure}

As we stated in the Sec.\ref{ssec::ska_lc_m_nu_025} 
we have assumed in our fiducial model a number of neutrino 
$N_\nu = 2$ which corresponds to an inverted hierarchy  
where the neutrinos are not all degenerate 
in mass (Fig.\ref{fig:neutrino_hyer}). By assuming that
$N_\nu$ is a continuous variable we can therefore 
compare and contrast models which assume a normal hierarchy and 
models that have an inverted hierarchy. A model with 
$N_\nu = 1$ would correspond to an normal hierarchy.

As we ran MCMC chains for LSS (SKA) data alone and CMB (Planck)
data alone we noted that
for a top-hat prior $0< N_\nu < 3$, which is what one would expect in 
a standard neutrino scenario without sterile neutrinos or any other exotic 
relativistic particle that could contribute significantly to the energy density of the Universe,
both of these experiments do not constrain $N_\nu$ significantly.
In fact as we can see from the relevant panels of Fig.\ref{fig::n2_matrix_ska} 
and Fig.\ref{fig::n2_matrix_planck},
the prior plays a bigger role in
the posterior than the experimental information itself if we consider each 
experiment independently. It would be more instructive if this were 
real data to widen the prior allowing more exotic scenarios to be allowed by the model.
However, when we combine both experiments we obtain a constraint which is slightly prior 
dependent but which is now mostly determined by data and this combination of data would yield 
a measurement of $ N_\nu = 2 \pm 0.5$ (1$\sigma$) and would reject the value
$N_\nu =1$ at just under 3$\sigma$.
We plot in Fig.\ref{fig::n2_both_ska} the expected posterior probability 
for the value of $N_\nu$
in this case.

\begin{figure}
\begin{center}
\centerline{\includegraphics[width=9.0cm,angle=0]{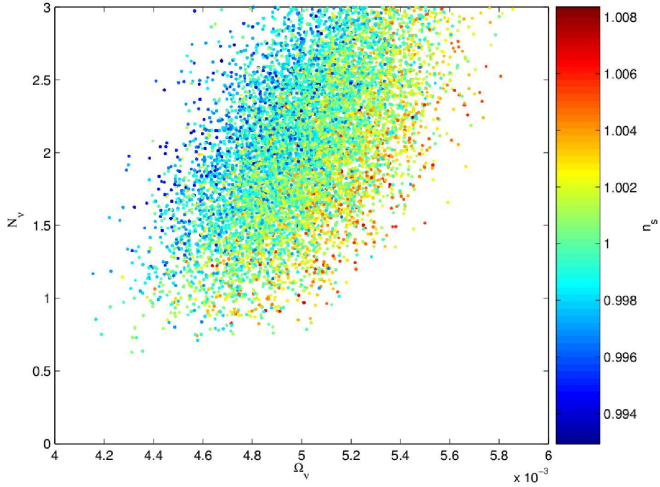}}
\caption[Confidence limits on neutrino hierarchies with SKA + Planck data.]{Confidence limits on neutrino hierarchies with LSS (SKA) and CMB (Planck) data. 
Illustration of the distribution of MCMC samples in the \{$N_\nu$, $\Omega_\nu$\} plane. 
The covariance around the fiducial \{2, 0.005\} point 
is such that it is possible for us to discriminate between a normal and inverted 
hierarchy. For higher values of a fiducial $\Omega_\nu$ 
the detection is clearer, 
and as we can see from Fig.\ref{fig::n1_omega} for lower values of 
a fiducial $\Omega_\nu$ the error becomes larger.}
\label{fig::n2_omega}
\end{center}
\end{figure}

This combination of $\Omega_\nu$ and $N_\nu$ would yield invaluable information
in order for us to be able to disentangle the origin of the neutrino mass and have a 
convincing theory which would explain the huge difference between the masses
of other particles and the neutrino mass. We plot in Fig.\ref{fig::n2_omega} the 
constraints of both of these parameters from a combination of LSS and CMB experiments.

\subsection{Indirectly probing neutrino hierarchies with cosmological data at $m_\nu \sim 0.05$ eV.}
\label{sec::indirect}

\begin{figure*}
\begin{center}
\centerline{
\includegraphics[width=8.0cm,angle=0]{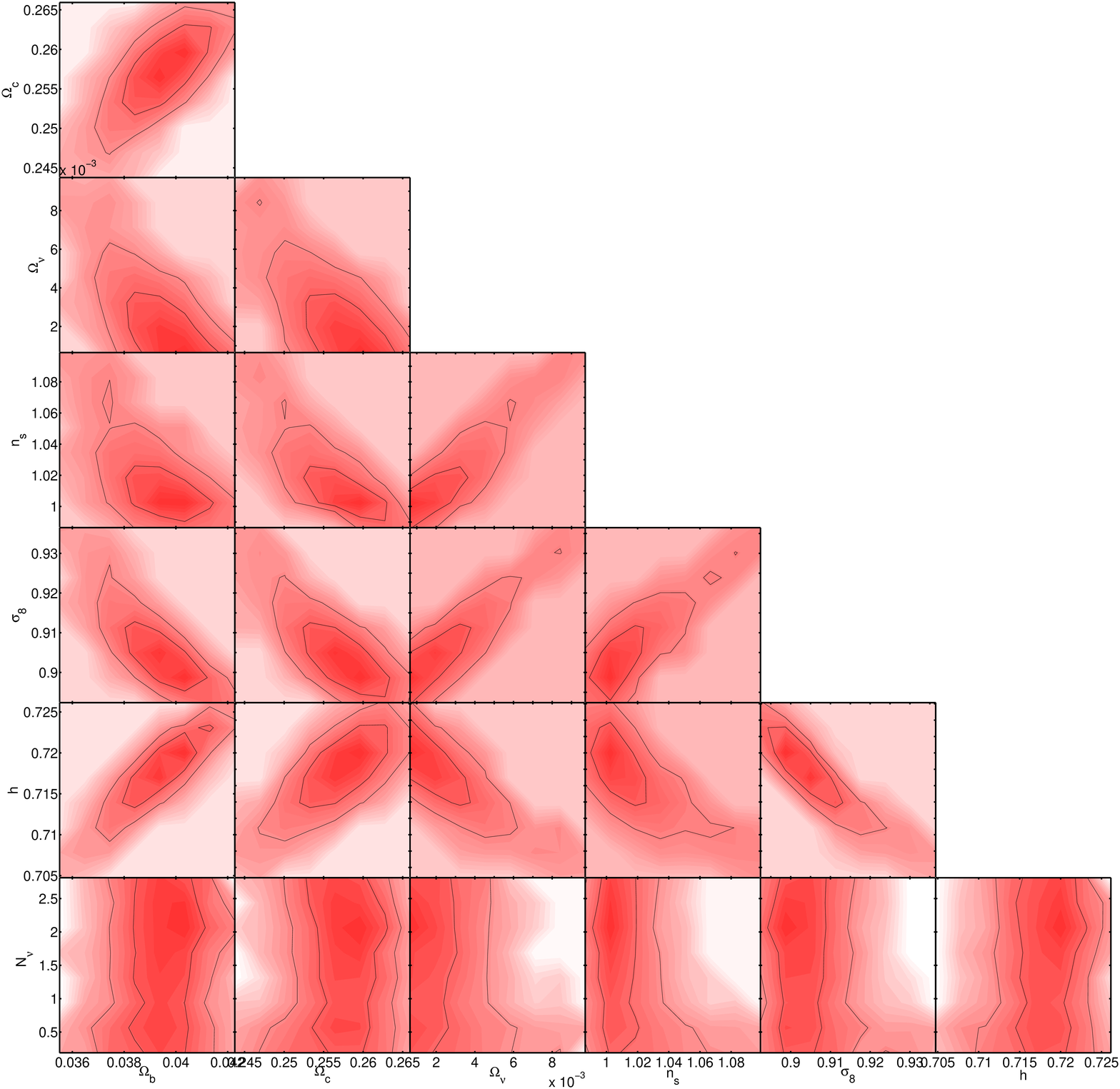}
\includegraphics[width=10.0cm,angle=0]{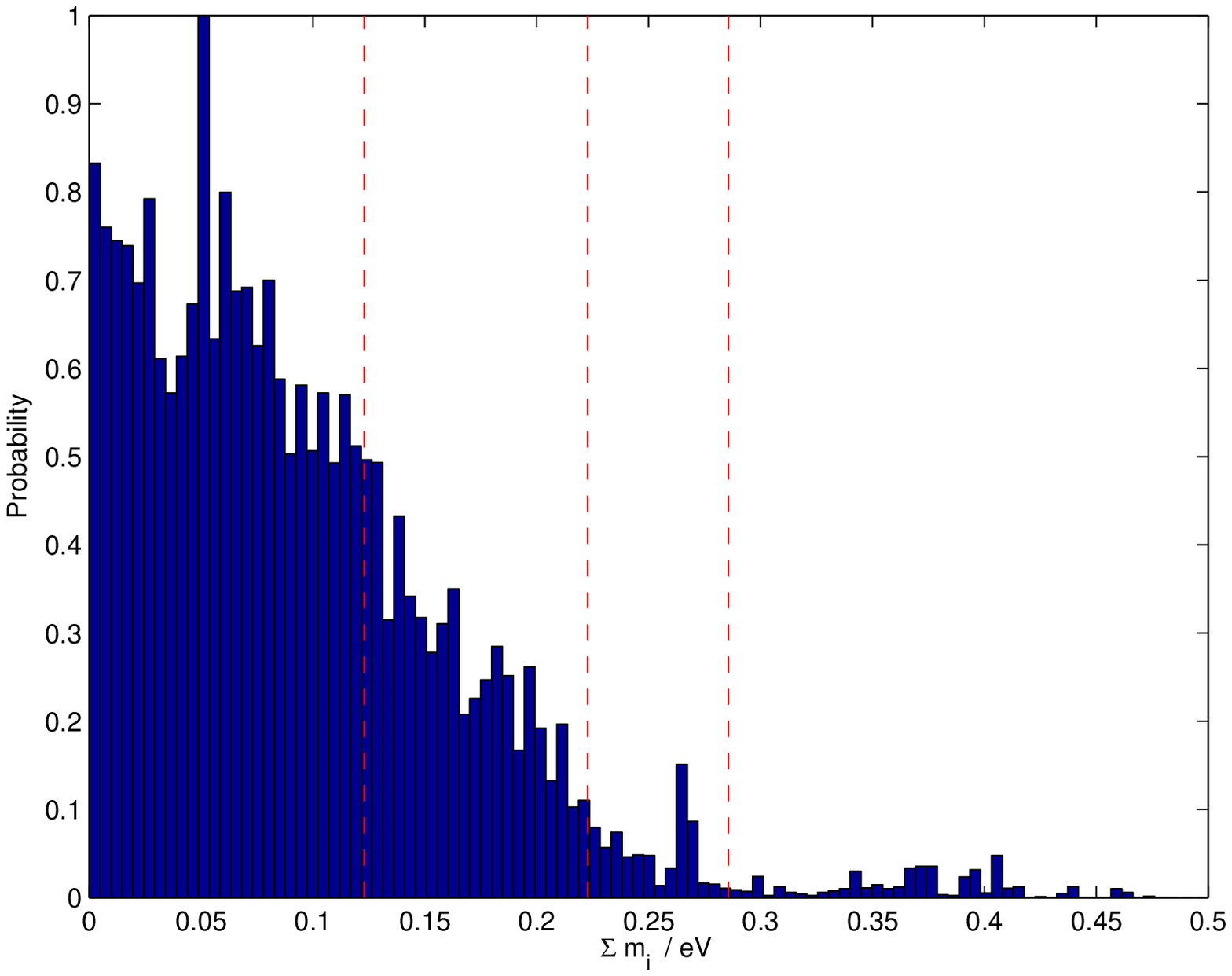}}

\caption[Confidence limits on neutrino hierarchies with SKA + Planck data.]{Illustration of MCMC samples assuming a total
energy density in relativistic particles $\Omega_\nu = 0.001$ and $N_\nu = 2$ which corresponds 
to an absolute neutrino mass scale of 0.05 eV. We consider LSS (SKA) simulated data alone.
The vertical red dashed lines in the right panel correspond to one,
two and three sigma confidence levels.
We can see that the parameters $N_\nu$ and $\Omega_\nu$ are not well measured,
even with exquisite data coming from future galaxy surveys. They alone will not provide evidence 
for a massive neutrino given the lower limit currently allowed by particle physics experiments.}
\label{fig::n1_ska}
\end{center}
\end{figure*} 

\begin{figure*}
\begin{center}

\begin{minipage}[c]{.75\textwidth} 
\centering 
\includegraphics[width=10.5cm,angle=0]{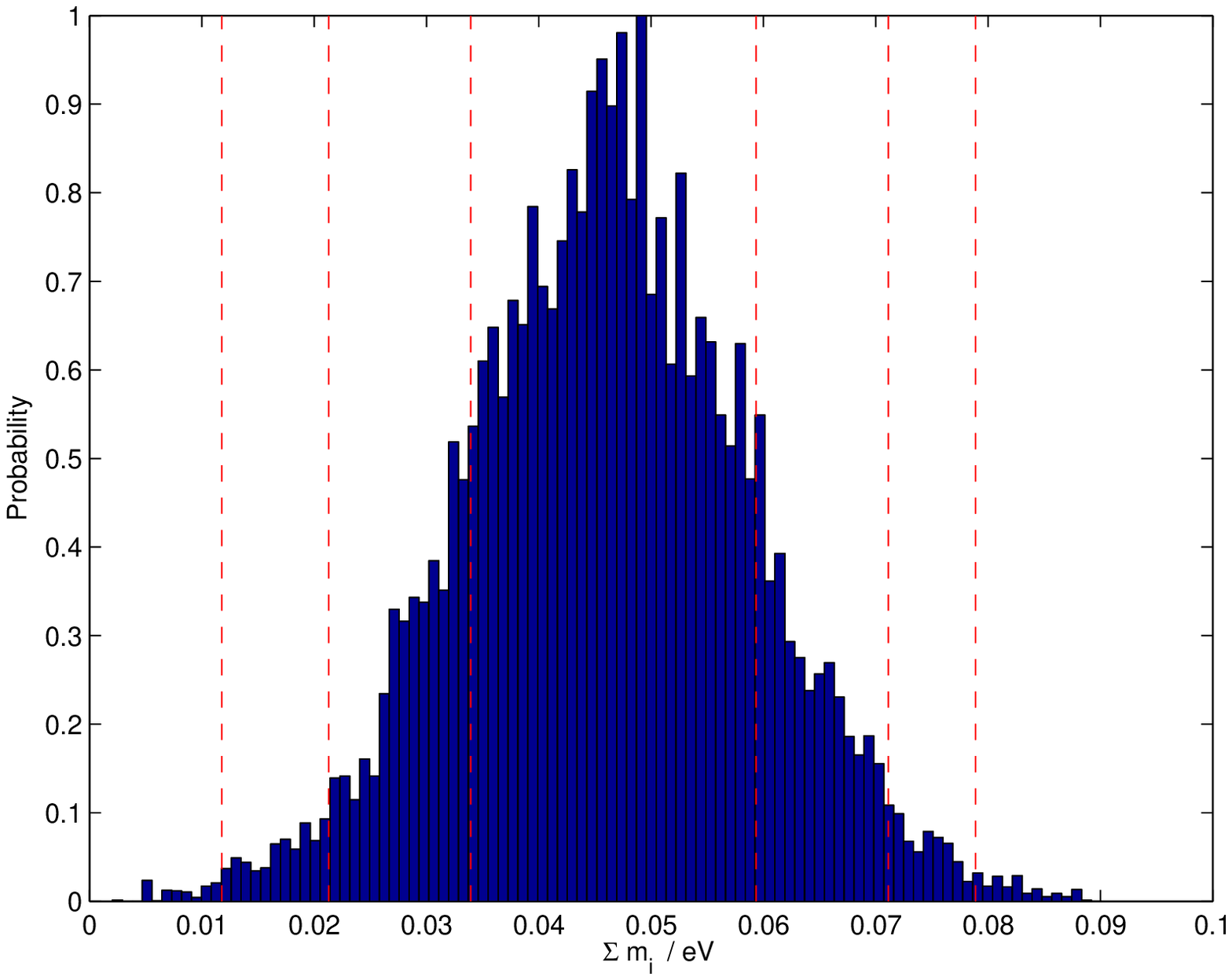}
\end{minipage}

\centerline{
\includegraphics[width=8.0cm,angle=0]{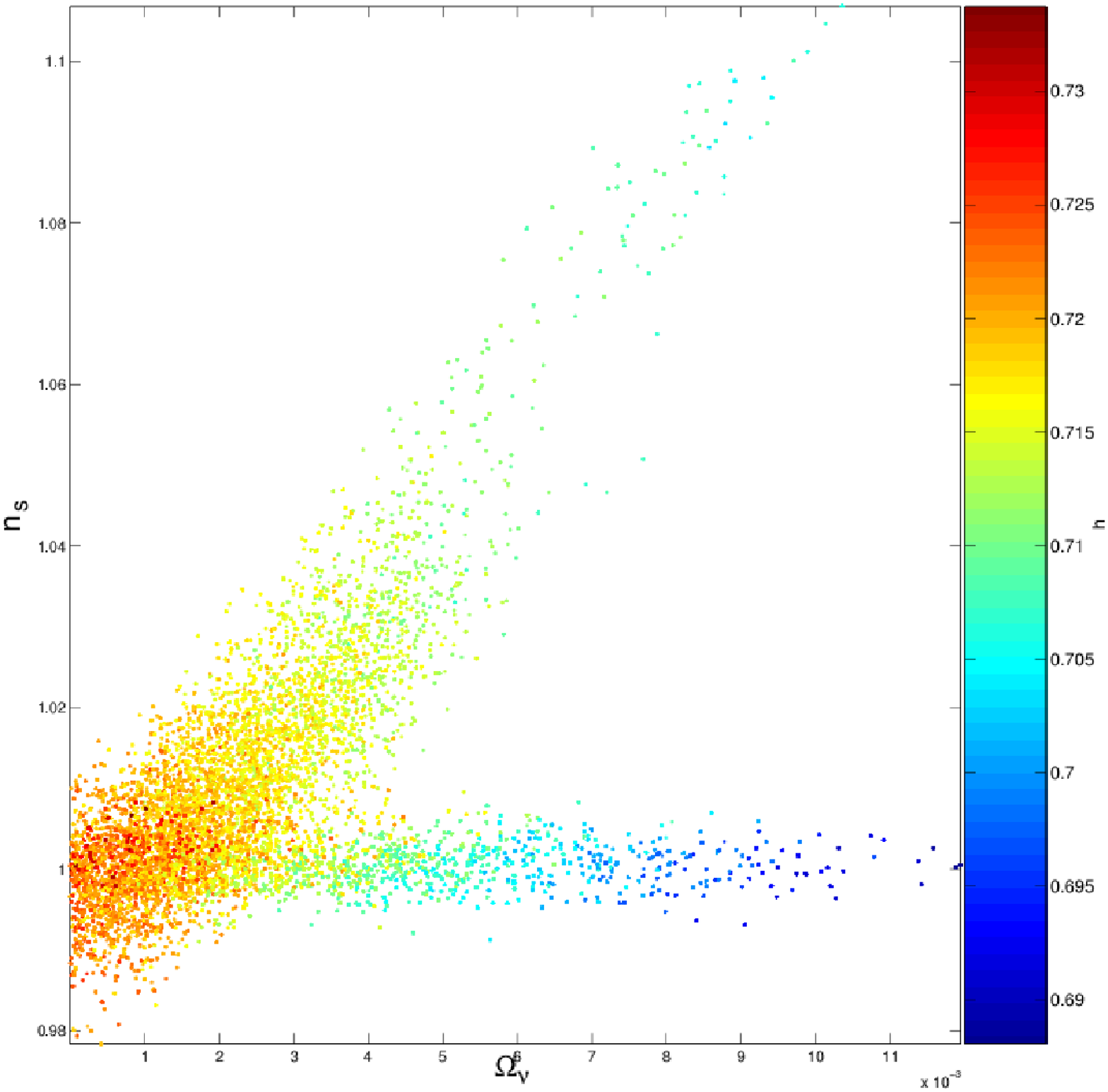}
\includegraphics[width=8.0cm,angle=0]{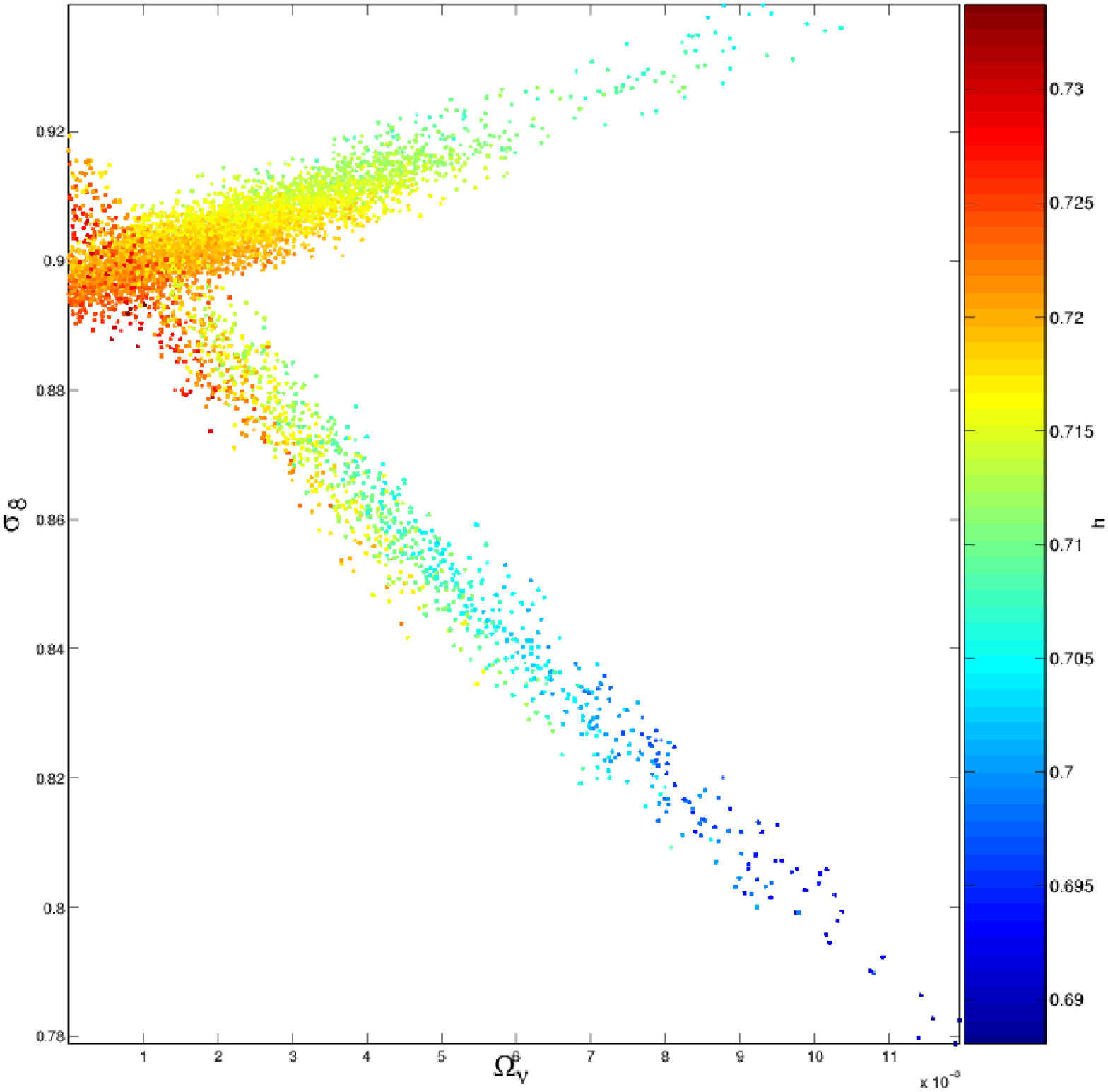}
}
\caption[Cosmic complementarity between LSS and CMB results ($m_\nu=0.05\, {\rm eV}$).]{We plot here the equivalent to Fig.\ref{fig::m_nu2_ska} and Fig.\ref{fig::degeneracies1_ska_cmb} 
using a fiducial cosmology with $\Omega_\nu = 0.001$, and considering
both LSS and CMB data, alone in the lower plots, and in combination in the upper panel.   
The vertical red dashed lines in the upper panel correspond to one,
two and three sigma confidence levels. We can see that 
LSS surveys will, alone, 
fail to measure a neutrino mass below 0.25 eV, and CMB surveys, alone, will also fail
to measure such a mass. However, as we can see from the upper pannel,
the combination of both surveys will be able to measure a neutrino mass as low
as 0.05 eV with an error of 0.015 eV.}
\label{fig::degeneracies2_ska_cmb}
\end{center}
\end{figure*}

We have shown that for a fiducial model where $m_\nu \sim 0.25 ~ \rm eV$ 
and $N_\nu =2$
there is a good prospect of a very significant detection of 
$m_\nu$ (Secs.\ref{ssec:ska_nu} and.\ref{ssec::ska_lc_m_nu_025}),  
and a reasonable prospect of a detection of $N_\nu$ 
with future SKA data (Sec.\ref{sec::direct_hi}). The problem is significantly 
harder if we wish to push the bounds to a neutrino mass of $m_\nu \sim 0.05$ 
eV which
is the limit one would like to attain given that it is the lower
limit of the parameter space allowed from particle physics
experiments.

We have run MCMC chains that assumed a fiducial model 
\{$\Omega_b$,$\Omega_c$,$\Omega_\nu$,$h$,$n_s$,$\sigma_8$, 
$N_{\nu}$\} = \{0.04,0.259,0.001,0.72,1.0,0.9,2\}, together with additional
nuisance parameters, in order to 
test whether the predictions we made hold and
to know whether future surveys will be able to push the limits
for $m_\nu$.

\begin{figure}
\begin{center}
\centerline{\includegraphics[width=9.0cm,angle=0]{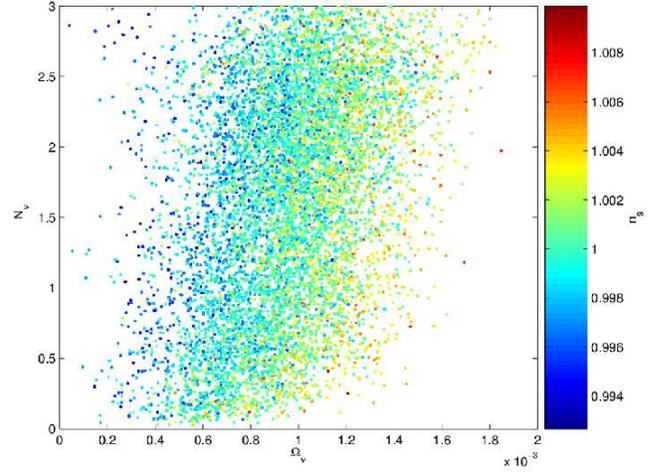}}
\caption[Confidence limits on neutrino hierarchies with SKA + Planck data.]{We
show here MCMC samples in the 2D plane \{$N_\nu$, $\Omega_\nu$\}: the fiducial model chosen
has a neutrino mass of 0.05 eV and 2 massive neutrino eigenstates, and both CMB and LSS data 
were considered. The covariance around 
the \{2, 0.001\} point is such that it is no longer possible for us to marginally detect
or rule out an inverted or normal hierarchy directly. However, as explained in
Sec.\ref{sec::indirect}, it is still possible to strongly prefer a hierarchy with the 
simple detection of the sum of the neutrino masses if $\sum m_i \sim$ 0.1 eV.}
\label{fig::n1_omega}
\end{center}
\end{figure}

We find that for an SKA survey alone the relevant parameters will not be 
measured accurately enough to provide strong bounds on the neutrino mass.
We plot in Fig.\ref{fig::n1_ska} the MCMC samples from such a
theoretical experiment. We find that a LSS (SKA) survey on its own will not
be able to find a signal either from the neutrino density parameter
or from the number density of massive neutrinos 
if $\sum m_i \sim 0.05$. It turns out that the
parameter degeneracy is too strong for there to be a clear detection.
A 3$\sigma$ upper limit for the neutrino mass would in this case be
$m_\nu \simeq 0.25\, {\rm eV}$ (95\% confidence).

However the addition of CMB data does improve significantly the 
estimate of $m_\nu $ at these low masses. As we can see from 
Fig.\ref{fig::degeneracies2_ska_cmb} it would be possible to measure the
neutrino mass accurately with $\sigma(m_\nu) \sim 0.015$ 
if we include both future LSS (SKA) and CMB (Planck) data. This would mean that
the entire parameter space that is currently still 
allowed by current cosmological
and laboratory experiments would be probed and a signal from a massive neutrino would
be detected.

It is possible for us to ask also whether it is possible to detect $N_\nu$
with this signal. We plot in Fig.\ref{fig::n1_omega} the posterior from 
the MCMC samples for the parameters $\Omega_\nu$ and $N_\nu$. Even though the 
mass density of neutrino can be measured accurately, $N_\nu$ is totally
unconstrained within the priors we have chosen. It would of course with real data make
sense to then widen the prior and obtain a sensible error for this parameter. However,
with a strong detection of the neutrino mass it is still possible within a certain 
range to obtain indirect conclusions with regards to the neutrino hierarchy.
If an inverted hierarchy is assume then we can simply assume that there are two massive neutrino and ignore the
mass of the third neutrino and consider it massless. In this case, given that there is a minimum mass 
splitting, and
we find a total sum of neutrino masses equal to less than twice this value then the
hierarchy cannot be inverted. This `reductio absurdum' method implies that
if $\sum m_i \ltsimeq 0.12\, eV$ then the 
hierarchy must be normal.

We conclude that a combined
data set comprising LSS (SKA) and CMB (Planck) would be able to detect the signal of a massive neutrino
any standard scenario. Furthermore, if the sum of neutrino masses is below 0.1 eV or above 0.25 eV there is a good possibility that
cosmological data will be able to prefer one hierarchy strongly. However if the sum of neutrino masses
is in the difficult range 0.1 - 0.25 eV then the data we have considered may not be enough 
for a strong preference to be found with regards to the hierarchy.

\section{Concluding remarks}

We have shown that future cosmological experiments will be able to 
detect a signal from a cosmological massive neutrino.
This signal would be encoded as a gentle curvature in
the galaxy power spectrum correlated with a shift of the
CMB temperature and polarisation peaks.
We argue that it will
be possible to remove any systematics present in data arising from 
the clustering properties of galaxies by the use of different
galaxy types in different redshift bins. Furthermore, data will be available 
from galaxies with different spatial clustering and hence any systematic effects 
will be assessed by comparing data sets with galaxies with different clustering
and redshift distributions.

We argue that for a clear detection to be made with confidence we need to be sure that any
prior used does not affect the results significantly. We find that future
cosmological data, both LSS (SKA) and CMB (Planck) data will be accurate and wealthy enough so that 
priors will play less of a role given that it will be able to measure several key parameters at
once including the matter density and the Hubble constant. We find as was already pointed out several times
in the literature, a strong complementarity between CMB and LSS data.

Although it is of utmost importance to measure properties of neutrinos
directly with particle physics experiments, there are several limitations 
to these techniques given that the mass scale of neutrinos is extremely small.
Therefore other probes from cosmological studies are key
to having a more complete picture of neutrino physics. Furthermore,
if we are to assess unusual theories such as scenarios where sterile particles
are present and have an energy density that influences 
cosmological data, then there needs to 
be detailed comparisons between the values 
drawn from direct particle physics detections and from cosmology.

We argue here that if the mass scale of the neutrino $m_\nu$ is very small 
we find ourselves necessarily 
in a non-degenerate scenario. However if this mass
scale is large enough, the spectrum of masses becomes quasi-degenerate and
most of the masses are almost equal. Current experiments cannot yet 
distinguish between these scenarios but may be on the verge of doing so. 
From a particle physics 
prospective the absolute mass scale of the neutrino is larger than 0.05 eV 
and smaller than 2 eV.

We have analysed to what extent neutrino parameters can be recovered
by combining future Planck data and a 
LSS survey, e.g. with the SKA, that would measure redshifts out to 
$z \sim 2$ over a large fraction of the sky. 
We conclude that this set of experiments would be sensitive over  
the entire mass range allowed by current particle physics experiments
down to 0.05 eV at the 3$\sigma$ level.
We find that even though the mass of the neutrino can be 
measured for the entire parameter space currently still allowed
by experiment, it is harder to measure the 
number density of massive neutrinos. We find that down to 0.25 eV
it is possible to measure the number density well enough
to distinguish between a scenario where one or two of the mass eigenstates
are massive and hence to be able to distinguish between a normal and
an inverted hierarchy.

We have ignored the effects of any varying equation of state of dark energy,
assuming that this value will mainly be determined by the angular scale of the
wiggles in the power spectrum. This information depends strongly on the 
value of $w$ and very weakly on the value of $\Omega_\nu$, hence providing 
orthogonal constraints as has already been seen in real data 
\citep{2006JCAP...06..019G}. Of course $w = -1$  could be the real answer.

We have also chosen to use a single value for $n_{s}$ so that the 
the running of the spectral index $n_{run}$ is set to zero. We argue that the degeneracy
that will degrade the neutrino sector constraints depends on
$n_s$ to first order and on $n_{run}$ to second order. 
The Planck
satellite is a mission designed to probe the shape of the initial power 
spectrum to unprecedented accuracy and 
hence will determine $\Delta n_{run} \sim 0.01$ (see Planck `blue book' 
\footnotemark). \footnotetext{http://www.rssd.esa.int/Planck}.

We have also assumed
the initial conditions to be adiabatic pertubations. 
Although it has been shown that isocurvature 
perturbations may generate degeneracies with neutrino parameters 
\citep{2006astro.ph.10597Z}, polarisation experiments from Planck will also
impose stringent limits on the amount of initial isocurvature. We therefore 
stress the importance of a high quality CMB mission as well as
a deep all-sky redshift survey in order to determine the neutrino parameters, 
because if this is not the case, limits from LSS alone would be also plagued by 
degeneracies.

We find that LSS (SKA) and CMB (Planck) 
data will constrain $\sum m_i$ down to 
0.05 eV with an error of around 0.015 eV. If $\sum m_i$
is indeed below 0.1 eV it will be possible to rule out an inverted hierarchy
indirectly given that having an inverted hierarchy with such small
$\sum m_i$ is in contradiction with atmospheric neutrino experiments. We therefore conclude that if the 
neutrino mass lies between 0.05 and 0.1 eV or above 0.25 eV it should be possible to 
distinguish clearly between the normal or inverted hierarchies.
However, it will not be possible to do so clearly if the mass lies between 0.1 and 0.25 eV.
These results may be modified  and improved, if a large range in data points can be used, and
specially if higher $k$ values can be included in the analysis by virtue of
improved modelling of non-linear fluctuations.

It is of great importance that cosmological experiments provide us with the same 
answers and parameters as particle physics experiments. Put another way,
it may eventually be possible to 
measure any exotic sterile particles that would contribute
to the cosmic background of relativistic particles and hence
influence $P(k)$ but which could not be detected by particle physics. 
We argue that it is likely to
be possible to constrain strongly any of 
these scenarios given that the data will be sensitive
down to $m_i \sim  0.05$ eV and $N_\nu \sim 1$.

\section*{Acknowledgements.}

We are very grateful for useful discussions with Steve Biller, Chris Blake, Sarah Bridle, Joanna Dunkley, Pedro Ferreira and Will Percival. We thank the Gemini Project and PPARC 
for a Studentship (FBA) and PPARC for a Senior Research Fellowship (SR). This research
has been undertaken as part of the SKA Design Study SKADS.

\bibliographystyle{./reference/mn2e.bst}

\bibliography{./reference/aamnem99,./reference/ref_data_base}

\begin{thebibliography}{}

\bibitem[\protect\citeauthoryear{{Abdalla}, {Blake} \& {Rawlings}}{{Abdalla}
  et~al.}{2006}]{ABR}
{Abdalla} F.~B.,  {Blake} C.~A.,    {Rawlings} S.,  2006, MNRAS to be submitted

\bibitem[\protect\citeauthoryear{{Abdalla} \& {Rawlings}}{{Abdalla} \&
  {Rawlings}}{2005}]{2005MNRAS.360...27A}
{Abdalla} F.~B.,  {Rawlings} S.,  2005, MNRAS, 360, 27

\bibitem[\protect\citeauthoryear{{Alcock} \& {Paczynski}}{{Alcock} \&
  {Paczynski}}{1979}]{1979Natur.281..358A}
{Alcock} C.,  {Paczynski} B.,  1979, Nat, 281, 358

\bibitem[\protect\citeauthoryear{{Ballinger}, {Peacock} \&
  {Heavens}}{{Ballinger} et~al.}{1996}]{1996MNRAS.282..877B}
{Ballinger} W.~E.,  {Peacock} J.~A.,    {Heavens} A.~F.,  1996, MNRAS, 282, 877

\bibitem[\protect\citeauthoryear{{Blake} \& {Glazebrook}}{{Blake} \&
  {Glazebrook}}{2003}]{2003ApJ...594..665B}
{Blake} C.,  {Glazebrook} K.,  2003, ApJ, 594, 665

\bibitem[\protect\citeauthoryear{{Bucher}, {Moodley} \& {Turok}}{{Bucher}
  et~al.}{2002}]{2002PhRvD..66b3528B}
{Bucher} M.,  {Moodley} K.,    {Turok} N.,  2002, Phys. Rev. D, 66, 023528

\bibitem[\protect\citeauthoryear{{Cole}, {Sanchez} \& {Wilkins}}{{Cole}
  et~al.}{2006}]{2006astro.ph.11178C}
{Cole} S.,  {Sanchez} A.~G.,    {Wilkins} S.,  2006, astro-ph/0611178

\bibitem[\protect\citeauthoryear{{Dunkley}, {Bucher}, {Ferreira}, {Moodley} \&
  {Skordis}}{{Dunkley} et~al.}{2005}]{2005MNRAS.356..925D}
{Dunkley} J.,  {Bucher} M.,  {Ferreira} P.~G.,  {Moodley} K.,    {Skordis} C.,
  2005, MNRAS, 356, 925

\bibitem[\protect\citeauthoryear{{Eisenstein}, {Hu} \& {Tegmark}}{{Eisenstein}
  et~al.}{1999}]{1999ApJ...518....2E}
{Eisenstein} D.~J.,  {Hu} W.,    {Tegmark} M.,  1999, ApJ, 518, 2

\bibitem[\protect\citeauthoryear{{Elgar{\o}y} \& {Lahav}}{{Elgar{\o}y} \&
  {Lahav}}{2003}]{2003JCAP...04..004E}
{Elgar{\o}y} {\O}.,  {Lahav} O.,  2003, Journal of Cosmology and Astro-Particle
  Physics, 4, 4

\bibitem[\protect\citeauthoryear{{Elgar{\o}y} \& {Lahav}}{{Elgar{\o}y} \&
  {Lahav}}{2005}]{2005NJPh....7...61E}
{Elgar{\o}y} {\O}.,  {Lahav} O.,  2005, New Journal of Physics, 7, 61

\bibitem[\protect\citeauthoryear{{Elgar{\o}y et al.}}{{Elgar{\o}y et
  al.}}{2002}]{2002PhRvL..89f1301E}
{Elgar{\o}y et al.} 2002, Physical Review Letters, 89, 061301

\bibitem[\protect\citeauthoryear{{Feldman}, {Kaiser} \& {Peacock}}{{Feldman}
  et~al.}{1994}]{1994ApJ...426...23F}
{Feldman} H.~A.,  {Kaiser} N.,    {Peacock} J.~A.,  1994, ApJ, 426, 23

\bibitem[\protect\citeauthoryear{{Fukuda et al.}}{{Fukuda et
  al.}}{1996}]{1996PhRvL..77.1683F}
{Fukuda et al.} 1996, Physical Review Letters, 77, 1683

\bibitem[\protect\citeauthoryear{{Fukugita}, {Ichikawa}, {Kawasaki} \&
  {Lahav}}{{Fukugita} et~al.}{2006}]{2006PhRvD..74b7302F}
{Fukugita} M.,  {Ichikawa} K.,  {Kawasaki} M.,    {Lahav} O.,  2006, Phys. Rev.
  D, 74, 027302

\bibitem[\protect\citeauthoryear{{Glazebrook} \& {Blake}}{{Glazebrook} \&
  {Blake}}{2005}]{2005ApJ...631....1G}
{Glazebrook} K.,  {Blake} C.,  2005, ApJ, 631, 1

\bibitem[\protect\citeauthoryear{{Goobar}, {Hannestad}, {M{\"o}rtsell} \&
  {Tu}}{{Goobar} et~al.}{2006}]{2006JCAP...06..019G}
{Goobar} A.,  {Hannestad} S.,  {M{\"o}rtsell} E.,    {Tu} H.,  2006, Journal of
  Cosmology and Astro-Particle Physics, 6, 19

\bibitem[\protect\citeauthoryear{{Guth}}{{Guth}}{1981}]{1981PhRvD..23..347G}
{Guth} A.~H.,  1981, Phys. Rev. D, 23, 347

\bibitem[\protect\citeauthoryear{{Hannestad}}{{Hannestad}}{2003}]{2003PhRvD..6%
7h5017H}
{Hannestad} S.,  2003, Phys. Rev. D, 67, 085017

\bibitem[\protect\citeauthoryear{{Hu}, {Eisenstein} \& {Tegmark}}{{Hu}
  et~al.}{1998}]{1998PhRvL..80.5255H}
{Hu} W.,  {Eisenstein} D.~J.,    {Tegmark} M.,  1998, Physical Review Letters,
  80, 5255

\bibitem[\protect\citeauthoryear{{Hu} \& {Haiman}}{{Hu} \&
  {Haiman}}{2003}]{2003PhRvD..68f3004H}
{Hu} W.,  {Haiman} Z.,  2003, Phys. Rev. D, 68, 063004

\bibitem[\protect\citeauthoryear{{Hu} \& {Tegmark}}{{Hu} \&
  {Tegmark}}{1999}]{1999ApJ...514L..65H}
{Hu} W.,  {Tegmark} M.,  1999, ApJ Lett., 514, L65

\bibitem[\protect\citeauthoryear{{Ichikawa}, {Fukugita} \&
  {Kawasaki}}{{Ichikawa} et~al.}{2005}]{2005PhRvD..71d3001I}
{Ichikawa} K.,  {Fukugita} M.,    {Kawasaki} M.,  2005, Phys. Rev. D, 71,
  043001

\bibitem[\protect\citeauthoryear{{Kaiser}}{{Kaiser}}{1987}]{1987MNRAS.227....1%
K}
{Kaiser} N.,  1987, MNRAS, 227, 1

\bibitem[\protect\citeauthoryear{{Kayser}}{{Kayser}}{2005}]{Kayser}
{Kayser} B.,  2005, SLAC Lecures 2004

\bibitem[\protect\citeauthoryear{{Lesgourgues} \& {Pastor}}{{Lesgourgues} \&
  {Pastor}}{2006}]{2006PhR...429..307L}
{Lesgourgues} J.,  {Pastor} S.,  2006, Phys. Rept., 429, 307

\bibitem[\protect\citeauthoryear{{Lewis} \& {Bridle}}{{Lewis} \&
  {Bridle}}{2002}]{2002PhRvD..66j3511L}
{Lewis} A.,  {Bridle} S.,  2002, Phys. Rev. D, 66, 103511

\bibitem[\protect\citeauthoryear{{Lewis}, {Challinor} \& {Lasenby}}{{Lewis}
  et~al.}{2000}]{2000ApJ...538..473L}
{Lewis} A.,  {Challinor} A.,    {Lasenby} A.,  2000, ApJ, 538, 473

\bibitem[\protect\citeauthoryear{{Maltoni}, {Schwetz}, {T{\'o}rtola} \&
  {Valle}}{{Maltoni} et~al.}{2004}]{2004NJPh....6..122M}
{Maltoni} M.,  {Schwetz} T.,  {T{\'o}rtola} M.,    {Valle} J.~W.~F.,  2004,
  hep-ph/0405172, New Journal of Physics, 6, 122

\bibitem[\protect\citeauthoryear{{Peacock} \& {Smith}}{{Peacock} \&
  {Smith}}{2000}]{2000MNRAS.318.1144P}
{Peacock} J.~A.,  {Smith} R.~E.,  2000, MNRAS, 318, 1144

\bibitem[\protect\citeauthoryear{{Peacock et al.}}{{Peacock et
  al.}}{2001}]{2001Natur.410..169P}
{Peacock et al.} 2001, nat, 410, 169

\bibitem[\protect\citeauthoryear{{Rawlings} \& {Abdalla}}{{Rawlings} \&
  {Abdalla}}{2006}]{RA}
{Rawlings} S.,  {Abdalla} F.~B.,  2006, Conference Procedings `On the Pathway
  to the SKA'

\bibitem[\protect\citeauthoryear{{Seljak}}{{Seljak}}{2000}]{2000MNRAS.318..203%
S}
{Seljak} U.,  2000, MNRAS, 318, 203

\bibitem[\protect\citeauthoryear{{Seljak} \& {Zaldarriaga}}{{Seljak} \&
  {Zaldarriaga}}{1996}]{1996ApJ...469..437S}
{Seljak} U.,  {Zaldarriaga} M.,  1996, ApJ, 469, 437

\bibitem[\protect\citeauthoryear{{Seljak et al.}}{{Seljak et
  al.}}{2005}]{2005PhRvD..71j3515S}
{Seljak et al.} 2005, Phys. Rev. D, 71, 103515

\bibitem[\protect\citeauthoryear{{Seo} \& {Eisenstein}}{{Seo} \&
  {Eisenstein}}{2003}]{2003ApJ...598..720S}
{Seo} H.-J.,  {Eisenstein} D.~J.,  2003, ApJ, 598, 720

\bibitem[\protect\citeauthoryear{{Spergel et al.}}{{Spergel et
  al.}}{2003}]{2003ApJS..148..175S}
{Spergel et al.} 2003, ApJS, 148, 175

\bibitem[\protect\citeauthoryear{{Spergel et al.}}{{Spergel et
  al.}}{2006}]{2006astro.ph..3449S}
{Spergel et al.} 2006, ArXiv astro-ph/0603449

\bibitem[\protect\citeauthoryear{{Springel}, {White}, {Jenkins}, {Frenk},
  {Yoshida}, {Gao}, {Navarro}, {Thacker}, {Croton}, {Helly}, {Peacock}, {Cole},
  {Thomas}, {Couchman}, {Evrard}, {Colberg} \& {Pearce}}{{Springel}
  et~al.}{2005}]{2005Natur.435..629S}
{Springel} V.,  {White} S.~D.~M.,  {Jenkins} A.,  {Frenk} C.~S.,  {Yoshida} N.,
   {Gao} L.,  {Navarro} J.,  {Thacker} R.,  {Croton} D.,  {Helly} J.,
  {Peacock} J.~A.,  {Cole} S.,  {Thomas} P.,  {Couchman} H.,  {Evrard} A.,
  {Colberg} J.,    {Pearce} F.,  2005, nat, 435, 629

\bibitem[\protect\citeauthoryear{{Tegmark et al.}}{{Tegmark et
  al.}}{2006}]{tegmark-2006-}
{Tegmark et al.} 2006, ArXiv astro-ph/0608632

\bibitem[\protect\citeauthoryear{{Verde et al.}}{{Verde et
  al.}}{2002}]{2002MNRAS.335..432V}
{Verde et al.} 2002, MNRAS, 335, 432

\bibitem[\protect\citeauthoryear{{Zunckel} \& {Ferreira}}{{Zunckel} \&
  {Ferreira}}{2006}]{2006astro.ph.10597Z}
{Zunckel} C.,  {Ferreira} P.~G.,  2006, astro-ph/0610597

\end{thebibliography}

%\end{thebibliography}

\end{document}